# Phase compositions of Si oxynitride: A comprehensive thermodynamic modeling study

Andrey Sarikov[1, 2, 3], Mariia Babiichuk[2], Mariia Voitovych[1], Sergii V. Mamykin[1]

[1] V. Lashkaryov Institute of Semiconductor Physics, National Academy of Sciences of Ukraine, 41 Nauky Avenue, 03028 Kyiv, Ukraine

[2] Educational Scientific Institute of High Technologies, Taras Shevchenko National University of Kyiv, 4-g Hlushkova Avenue, 03022 Kyiv, Ukraine

[3] National Technical University of Ukraine "Igor Sikorsky Kyiv Polytechnic Institute", 37 Beresteiskyi Avenue, 03056 Kyiv, Ukraine

Corresponding author's email: sarikov@isp.kiev.ua

**Abstract**

In this work, a comprehensive thermodynamic description of the formation of phases in amorphous $SiO_xN_y$ ($0 \leq x \leq 2$, $0 \leq y \leq 4/3$) films as a function of the film chemical composition and fabrication temperature is presented. Two possible phase states, namely homogeneous single-phase Si oxynitride and two-phase mixture of Si oxide and Si nitride phases, are analyzed in terms of the lowest values of the Gibbs free energy. The characters of the modification of the phase composition upon changing the $SiO_xN_y$ stoichiometry indices and fabrication temperature are determined. Moreover, characteristic features of the chemical composition of the Si oxide and Si nitride phases in two-phase Si oxynitrides are revealed. Particular attention is given to the thermodynamic origin of the boundaries between different phase states of the $SiO_xN_y$ films and to determining the critical stoichiometry values separating them. A comprehensive phase diagram indicating single- and two-phase $SiO_xN_y$ regions as functions of the fabrication temperature is constructed. The peculiarities of the transitions between different $SiO_xN_y$ phase states, along with underlying mechanisms, are discussed with reference to this phase diagram. The obtained results extend our understanding of the fundamental properties of Si oxynitride films and their dependence on the film fabrication conditions, and may be used for tailoring $SiO_xN_y$ characteristics for practical applications.

## 1. Introduction

Amorphous stoichiometric and non-stoichiometric (Si-rich) Si oxynitride ($SiO_xN_y$, $0 \leq x \leq 2$, $0 \leq y \leq 4/3$) is a composite material, whose mechanical, optical, electrical, and luminescence properties can be varied in wide ranges by varying its chemical composition (stoichiometry indices $x$ and $y$). Various aspects of the technology, properties and applications of this material are disclosed in particular in a large number of review and original papers published during the last years [1-16]. $SiO_xN_y$ films find multiple applications in modern nano- and optoelectronics such as for thin-film passivation and encapsulation layers [3, 13, 15], anti-reflection coatings in gradient-index optics and waveguide structures [7, 16], dielectric layers in resistive random access memories [17], buffer layers for organic photodetecting devices reducing dark currents and enhancing the device thermal stability [18], solar cell emitter layers [19], and protective coatings for transparent polymer-based devices slowing down oxygen removal and suppressing peroxide-induced defect formation thus extending the polymer film durability [20, 21]. The adjustable electronic structure as well as presence of sensitizers, such as Si nanoclusters and agglomerates, and defect levels within the band gap make the $SiO_xN_y$ films perspective for light-emitting elements with tunable emission wavelength [5, 9, 12, 22, 23]. In biomedicine, Si oxynitride films are introduced as titanium implants coatings in dentistry and orthopedics [24-26] as well as in biomedical sensing [27, 28]. Such films combine biocompatibility with chemical stability presumably due to presence of Si–OH groups in the near-surface layers, which act as attachment sites for calcium and phosphorus atoms from biological fluids thereby facilitating growth of a critically important bonding layer between the implant and the bone tissue [25, 26]. Recently, use of reflection coatings based on Si oxynitride layers for rapidly developing gravitational wave detectors has been proposed [29].

For creating the required $SiO_xN_y$ material, the mechanisms and peculiarities of the formation of its properties must be clearly identified. It should be noticed that the properties of the $SiO_xN_y$ films are defined not solely by their stoichiometry but also by microstructure and phase composition, which may be non-equivalent for the films with equal $x$ and $y$ values but grown at different technological conditions. Two principal models are traditionally used to describe the phase structure of the Si oxynitrides, namely the random mixture model (RMM) and the random bonding model (RBM). According to the former model, $SiO_xN_y$ consists of a mixture of $SiO_2$-like and $Si_3N_4$-like phases [30]. Si-rich oxynitrides may additionally contain amorphous Si clusters [31-33] often accompanied by Si–H bond formation [33-35]. In contrast, the RBM assumes chemical homogeneity of the $SiO_xN_y$ microstructure with O and N atoms statistically distributed among five types of tetrahedral complexes $Si–O_aN_{4–a}$ ($a = 0..4$) with a central Si atom [36, 37]. For Si-rich compositions, presence of Si–Si bonds must additionally be taken into account leading to a description of the local structure by $Si–O_aN_bSi_{4–a–b}$ tetrahedral configurations with $a$, $b = 0..4$ and $a + b \leq 4$ [30]. Moreover, an intermediary model, in which $SiO_2$ and $Si_3N_4$ inclusions are distributed within a homogeneous $SiO_xN_y$ matrix, was proposed [38, 39]. Based on a detailed analysis of infrared (IR) absorption spectra of $SiO_xN_y$ films grown by plasma-enhanced chemical vapor deposition (PECVD) as well as of vast variety of literature data, we demonstrated the regularities of the dependence of the films phase composition on their chemical composition and fabrication temperature [40]. Namely, decrease in the relative Si content in the films (increase in the stoichiometry indices *x* and *y*) as well as raise of the fabrication temperature lead to homogeneous Si oxynitrides obeying the random bonding model. On the other hand, the films with intermixed generally non-stoichiometric Si oxide and Si nitride phases are obtained at smaller stoichiometry indices and/or low temperatures. The structures of Si oxynitride films obtained by different technologies corresponding to either of the mentioned models were confirmed experimentally using Rutherford backscattering (RBS), Fourier-transform infrared (FTIR) spectroscopy, X-ray photoelectron spectroscopy (XPS), X-ray near edge spectroscopy (XANES), transmission electron microscopy (TEM), and extended X-ray absorption fine structure (EXAFS) spectroscopy [3, 9, 11, 30, 31, 36-43].

The existence of different structural states of $SiO_xN_y$ raises a fundamental question about the thermodynamic origin of the preference of either a homogeneous amorphous Si–O–N network or a separated into Si oxide and Si nitride phases structure. To the best knowledge of us, no attempts have been undertaken so far to address this issue comprehensively. For Si oxides, a thermodynamic theory covering all the major aspects of formation of phases was presented in [44]. In [45, 46], computational thermodynamics with the parameters for gaseous precursors and condensed phases from standard databases was applied to study the composition and phase composition of $SiO_xN_y$ obtained by chemical vapor deposition, as functions of the precursor types and ratios. At this, the Si oxynitride phase composition was considered in terms of the Si, $SiO_2$, $Si_3N_4$ and $Si_2O_2N$ phases. In our previous study [40], a theoretical model was developed relating a particular phase composition of a $SiO_xN_y$ film to its relative thermodynamic stability expressed in terms of the minimum value of the Gibbs free energy. This model enabled describing the peculiarities of transitions between the homogeneous and phase-separated into Si oxide and Si nitride states of the $SiO_xN_y$ films upon changing the film chemical composition and fabrication temperature. Subsequent application of this model provided a basis for constructing phase diagram of Si oxynitride [47] to indicate stoichiometry regions corresponding to single-phase and two-phase $SiO_xN_y$ structures. A consistency of the calculated boundaries between different phase states with experimentally determined $SiO_xN_y$ phase compositions was demonstrated.

Nevertheless, a comprehensive description of the $SiO_xN_y$ phase compositions and transitions between single-phase and two-phase states over the entire accessible stoichiometry range is still lacking. In particular, thermodynamic mechanisms responsible for the formation of particular phase states are not properly discussed. Moreover, the critical values of the stoichiometry indices *x* and *y* delineating the homogeneous and separated into Si oxide and Si nitride states, and their dependence on the $SiO_xN_y$ film fabrication temperature require deep systematic analysis. Quantitative determination of the boundaries between different phase compositions would provide not only better interpretation of experimental structural data but also a practical means for controlling the properties of $SiO_xN_y$ films during fabrication.

In this work, we present a comprehensive thermodynamic description of the formation of phase compositions of amorphous $SiO_xN_y$ over the stoichiometry range $0 \le x \le 2$ and $0 \le y \le 4/3$. Thermodynamic equilibrium between the competing homogeneous and separated two-phase states is analyzed as a function of the $SiO_xN_y$ chemical composition and temperature. Particular attention is given to the thermodynamic origin of the boundaries between the mentioned phase states and to determining the critical $x$ and $y$ values separating them. Moreover, characteristic features of the chemical composition of the Si oxide and Si nitride phases in two-phase $SiO_xN_y$ systems are described. A comprehensive phase diagram is constructed indicating the regions of the stability of single- and two-phase $SiO_xN_y$ compositions. The peculiarities of transitions between different phase states are discussed with reference to this phase diagram. The obtained results provide a quantitative framework for predicting and controlling phase composition of $SiO_xN_y$ films and form a basis for establishing relationships between the thermodynamic state, microstructure, and functional properties of such films.

## 2. Theoretical model

In this work, formation of phases in $SiO_xN_y$ films as a function of the film stoichiometry (values of $x$ and $y$) and fabrication temperature is studied by thermodynamic modeling. The model proposed in [40] is used for the study. Namely, we analyze the dependence of the Gibbs free energy of the $SiO_xN_y$ films on their phase and chemical composition and temperature. Single-phase films with homogeneous distribution of O and N atoms as well as two-phase films consisting of regions of generally non-stoichiometric Si oxide and nitride phases are considered. Preference of a certain type of the $SiO_xN_y$ phase composition is determined according to the lowest-free-energy criterion.

We considered $SiO_xN_y$ films as formed by tetrahedral structural units $Si–O_aN_bSi_{4–a–b}$ ($0 \le a + b \le 4$) with a central Si atom in a case of a homogeneous single-phase system, or $Si–O_aSi_{4–a}$ (Si oxide regions) and $Si–N_bSi_{4–b}$ (Si nitride regions) with $0 \le a, b \le 4$ in a case of a two-phase system. The distribution of the units in the single-phase system, as well as in each of the phases in the two-phase system was described by the random bonding model. Each unit was assigned a penalty energy value $\Delta_{ab}$, $\Delta_a$ or $\Delta_b$ characterizing non-equivalence in energy of the units with different oxidation degrees of the central Si atom. The penalty energy concept was adopted from the description of energetics of non-stoichiometric Si oxides [48, 49], where the penalty energy $\Delta_a$ of a $Si–O_aSi_{4–a}$ structural unit was introduced as the difference between the ab initio calculated unit energy and the sum of $a$ Si–O and $4–a$ Si–Si bond energies assumed to be independent on $a$. The values $\Delta_a$ for Si oxides were found to be $\Delta_0 = \Delta_4 = 0$, $\Delta_1 = 0.5$ eV, $\Delta_2 = 0.51$ eV, and $\Delta_3 = 0.22$ eV [49].

The Gibbs free energy of a homogeneous $SiO_xN_y$, $G_{hom}(x, y, T)$, consists of the total penalty energy of all the structural units $Si–O_aN_bSi_{4–a–b}$, $G_{oxn}^{penalty}(x, y)$, and the contribution of the configuration entropy $\Omega_{oxn}(x, y)$ associated with the number of possible arrangements of oxygen and nitrogen atoms between Si atoms [40]:

$$G_{hom}(x, y, T) = G_{oxn}^{penalty}(x, y) - \Omega_{oxn}(x, y) \cdot T \qquad (1)$$

where $T$ is the temperature.

According to the random bonding model, $G_{oxn}^{penalty}$ can be expressed as follows:

$$G_{oxn}^{penalty}(x, y) = N_{Si} \sum_{a=0}^{4} \sum_{b=0}^{4-a} \frac{4!}{a!b!(4-a-b)!} \left(\frac{x}{2}\right)^a \left(\frac{3y}{4}\right)^b \left(1 - \frac{x}{2} - \frac{3y}{4}\right)^{4-a-b} \Delta_{ab} \qquad (2)$$

where $N_{Si}$ is the total number of Si atoms in the considered Si oxynitride. Since no penalty energy values for the structural units containing nitrogen atoms are available in the literature, but there are indications of their proximity for the case of O and N atoms [50], all the calculations were carried out assuming $\Delta_{ab} = \Delta_{a+b}$.

The configuration entropy $\Omega_{oxn}(x, y)$ within the random bonding model is expressed as follows:

$$\Omega_{oxn}(x, y) = kN_{Si} \left(2 \ln \frac{4}{4-2x-3y} - x \ln \frac{2x}{4-2x-3y} - \frac{3y}{2} \ln \frac{3y}{4-2x-3y}\right) \qquad (3)$$

where $k = 8.6 \times 10^{-5}$ eV/K is the Boltzmann constant.

In the two-phase $SiO_xN_y$ cases, the stoichiometries of the Si oxide and Si nitride phases were described by introducing the coefficient α ($0 \leq \alpha \leq 1$), indicating the portion of Si atoms in the Si oxide phase. Then the stoichiometry indices of the Si oxide ($\acute{x}$) and nitride ($\acute{y}$) phases are expressed as follows:

$$\acute{x} = \frac{x}{\alpha}, \qquad \acute{y} = \frac{y}{1-\alpha} \qquad (4)$$

The value of α changes upon redistribution of Si atoms between the Si oxide and nitride phases to minimize the Gibbs free energy of the two-phase Si oxynitride. The minimum possible value of α, $\min \alpha = \frac{x}{2}$, corresponds to the maximum value $\acute{x} = 2$, i.e. formation of stoichiometric $SiO_2$ phase, all the rest Si atoms entering the Si nitride phase. The maximum possible value of α, $\max \alpha = 1 - \frac{3}{4}y$, corresponds to the maximum value $\acute{y} = \frac{4}{3}$, i.e. formation of stoichiometric $Si_3N_4$ phase with all the rest Si composing the Si oxide phase.

The Gibbs free energy of a heterogeneous two-phase system, $G_{het}(x, y, T, \alpha)$, is calculated as a sum of the contributions from the Si oxide and nitride phases as follows:

$$G_{het}(x, y, T, \alpha) = G_{oxide}^{penalty}(x, \alpha) - \Omega_{oxide}(x, \alpha) \cdot T + G_{nitride}^{penalty}(y, \alpha) - \Omega_{nitride}(y, \alpha) \cdot T \tag{5}$$

The total penalty energy $G_{oxide}^{penalty}$ and the configuration entropy $\Omega_{oxide}$ of the Si oxide are expressed as follows:

$$G_{oxide}^{penalty}(x, \alpha) = \alpha N_{Si} \sum_{a=0}^{4} \frac{4!}{a!(4-a)!} \left(\frac{\acute{x}}{2}\right)^{a} \left(1 - \frac{\acute{x}}{2}\right)^{4-a} \Delta_a \tag{6a}$$

$$\Omega_{oxide}(x, \alpha) = \alpha k N_{Si} \left(2 \ln \frac{2}{2-\acute{x}} - \acute{x} \ln \frac{\acute{x}}{2-\acute{x}}\right) \tag{6b}$$

Similar relations can be obtained for the Si nitride phase:

$$G_{nitride}^{penalty}(y, \alpha) = (1-\alpha) N_{Si} \sum_{b=0}^{4} \frac{4!}{b!(4-b)!} \left(\frac{3\acute{y}}{4}\right)^{b} \left(1 - \frac{3\acute{y}}{4}\right)^{4-b} \Delta_b \tag{7a}$$

$$\Omega_{nitride}(y, \alpha) = (1-\alpha) k N_{Si} \left(2 \ln \frac{4}{4-3\acute{y}} - \frac{3\acute{y}}{2} \ln \frac{3\acute{y}}{4-3\acute{y}}\right) \tag{7b}$$

The absolute minimum of the expression (5) with respect to α, $\min_{\alpha} G_{het}(x, y, T, \alpha)$, defines the compositions of the Si oxide and Si nitride phases in two-phase $SiO_xN_y$ with the stoichiometry indices $x$ and $y$ at the temperature $T$. Single- or two-phase $SiO_xN_y$ system is formed depending on which of the two expressions $G_{hom}(x, y, T) < \min_{\alpha} G_{het}(x, y, T, \alpha)$ or $G_{hom}(x, y, T) > \min_{\alpha} G_{het}(x, y, T, \alpha)$ holds.

## 3. Results

### *3.1. Phase compositions of two-phase Si oxynitrides*

In this section, the predicted peculiarities of the phase compositions of two-phase Si oxynitrides are discussed.

According to the theoretical model (see expressions (4) to (7)), the parameter α indicates the distribution of Si atoms between generally non-stoichiometric Si oxide and Si nitride phases in a two-phase Si oxynitride. Figure 1 presents calculated exemplary dependences of the Gibbs free energy of two-phase $SiO_xN_y$ films ($G_{hom} > \min_{\alpha} G_{het}(\alpha)$) with stoichiometries in a wide range and for different fabrication temperatures on the parameter α. These dependences illustrate a characteristic feature of the Gibbs free energy of a two-phase Si oxynitride, namely existence of two minima located at close to the lowest and the highest possible values of α. As described in the previous section, the position of the absolute minimum defines the chemical compositions of the phases in the two-phase Si oxynitride. The left absolute minimum ($\alpha_{min} \to \min \alpha$) corresponds to $SiO_xN_y$ with close to stoichiometric Si oxide ($SiO_2$) phase, the rest of the Si atoms forming non-stoichiometric Si nitride phase. Similarly, the right absolute minimum ($\alpha_{min} \to \max \alpha$) indicates formation of close to stoichiometric Si nitride ($Si_3N_4$) phase with the rest of the Si atoms making up non-stoichiometric Si oxide phase. Referring to the exemplary cases depicted in Figure 1, the following phase states of the $SiO_xN_y$ films are predicted:

- $SiO_{0.8}N_{0.3}$ fabricated at 400°C – a two-phase system with close to stoichiometric $SiO_2$ and close to $SiN_{0.5}$ phases (Figure 1(a));
- $SiO_{0.5}N_{0.4}$ fabricated at 400°C – a two-phase system with close to stoichiometric $Si_3N_4$ and close to $SiO_{0.71}$ phases (Figure 1(b));
- $SiO_{1.2}N_{0.1}$ fabricated at 1000°C – a two-phase system with close to stoichiometric $SiO_2$ and close to $SiN_{0.25}$ phases (Figure 1(c)); and
- $SiO_{0.1}N_{0.8}$ fabricated at 1000°C – a two-phase system with close to stoichiometric $Si_3N_4$ and close to $SiO_{0.25}$ phases (Figure 1(d)).

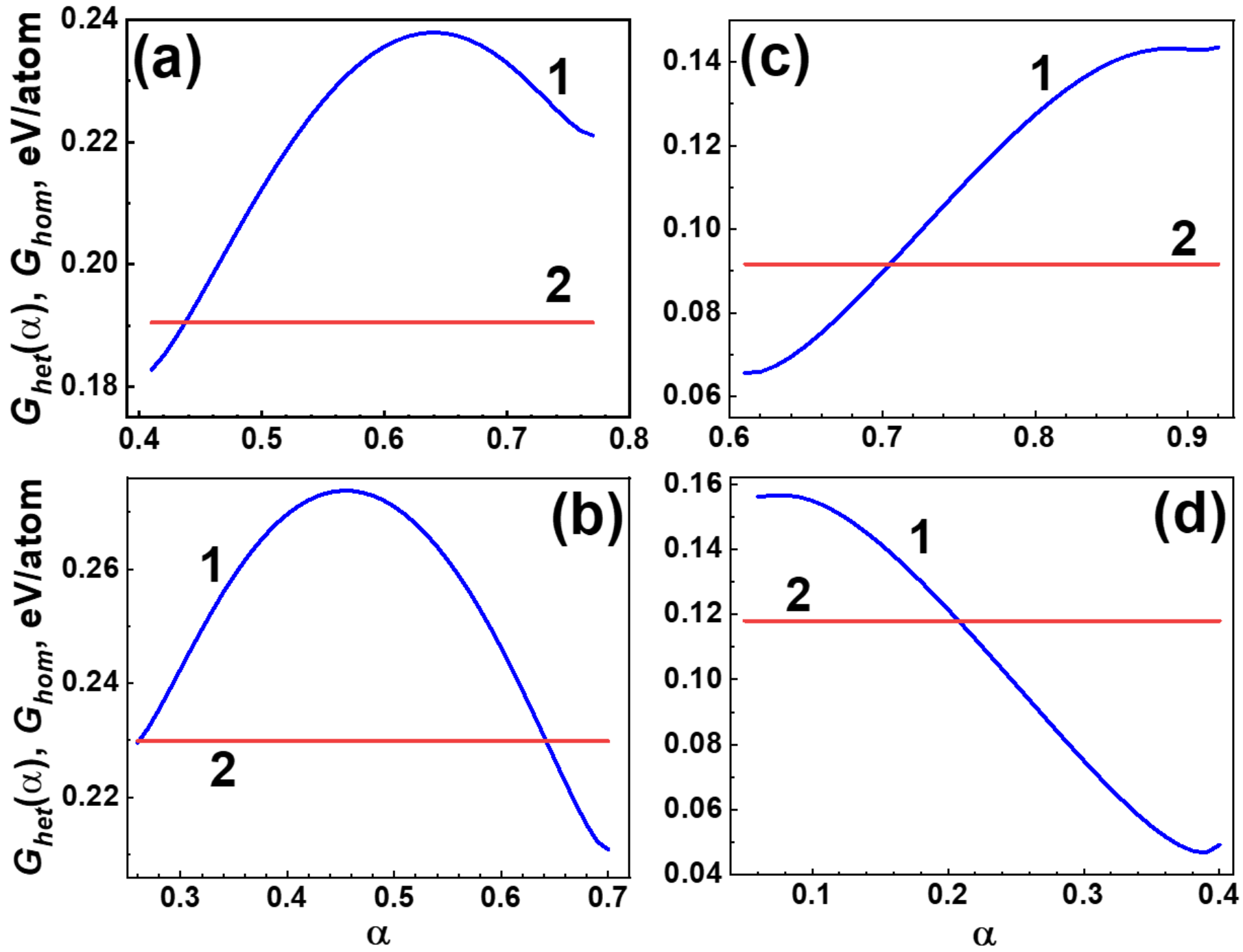

**Figure 1.** Gibbs free energy of two-phase (1) and single-phase (2) Si oxynitride: (a) – $SiO_{0.8}N_{0.3}$, $T$ = 400°C, (b) – $SiO_{0.5}N_{0.4}$, $T$ = 400°C, (c) – $SiO_{1.2}N_{0.1}$, $T$ = 1000°C, and (d) – $SiO_{0.1}N_{0.8}$, $T$ = 1000°C.

Hence, the results of the investigations of the dependence of the Gibbs free energy on the parameter α provide the following property of Si oxynitrides separated into Si oxide and Si nitride phases: Whenever a two-phase Si oxynitride is formed one of its phases (either Si oxide or Si nitride) necessarily has the composition close to the stoichiometric one, while the other phase comprises all the rest of the Si atoms and remains generally non-stoichiometric.

### *3.2. Influence of fabrication temperature on $SiO_xN_y$ phase composition*

In this section, the effect of fabrication temperature on the phase composition of Si oxynitride films is investigated. This effect is illustrated by Figure 2, where the dependences of the Gibbs free energy on the parameter α at different temperatures are shown for two exemplary $SiO_xN_y$ cases corresponding to different phase compositions in a two-phase state, namely the one with close to stoichiometric $SiO_2$ phase ($SiO_{0.8}N_{0.3}$, panels (a) to (c)) and the other one with close to stoichiometric $Si_3N_4$ phase ($SiO_{0.2}N_{0.8}$, panels (d) to (f)). As can be seen from this figure, the Gibbs free energy of the single-phase system in both cases decreases relative to the energy of the two-phase system upon increasing the temperature. Calculations by expressions (1) and (5) show that this process takes place due to faster decrease with temperature in the entropy contribution to the Gibbs free energy of a single-phase Si oxynitride as compared to the respective two-phase system. This result is repeated for all the investigated $SiO_xN_y$ chemical compositions. Therefore, a conclusion is drawn that increase in the $SiO_xN_y$ deposition temperature induces transition from a two-phase state with separated Si oxide and Si nitride phases to homogeneous single-phase Si oxynitride, and decrease in this temperature favors reverse transition. This conclusion coincides with that made based on the IR absorption spectroscopy results and analysis of literature data carried out in [40], as was mentioned in the Introduction section. Moreover, the two-phase/single-phase transition temperature is dependent on the $SiO_xN_y$ chemical composition, which will be discussed in Section 4 with reference to the phase diagram of Si oxynitride systems.

It should be also noted that increase in the fabrication temperature of a Si oxynitride film induces shifts of the minima of the Gibbs free energy of the two-phase systems away from the maximum and minimum possible values of α, which corresponds to an increase in the concentrations of excess Si in the Si oxide and Si nitride phases (decrease in the stoichiometry indices $\acute{x}$ and $\acute{y}$, see expression (4)). A similar effect was described for non-stoichiometric Si oxides [51], caused by the increasing role of the entropy contribution to the Gibbs free energy at higher temperatures. At the same time, as can be further seen from Figure 2, the temperature changes do not alter the qualitative relationship between the Gibbs free energy values of the two-phase systems in the left and right minima. This finding can be understood by analyzing the expression (5), where only entropy-related contributions to the Gibbs free energy of a two-phase system are temperature-dependent and scale proportionally to it. This means that no transformation from a close-to-$SiO_2$-containing two-phase Si oxynitride to a close-to-$Si_3N_4$-containing one or vice versa is possible for a Si oxynitride film with a fixed chemical composition upon varying the film fabrication temperature.

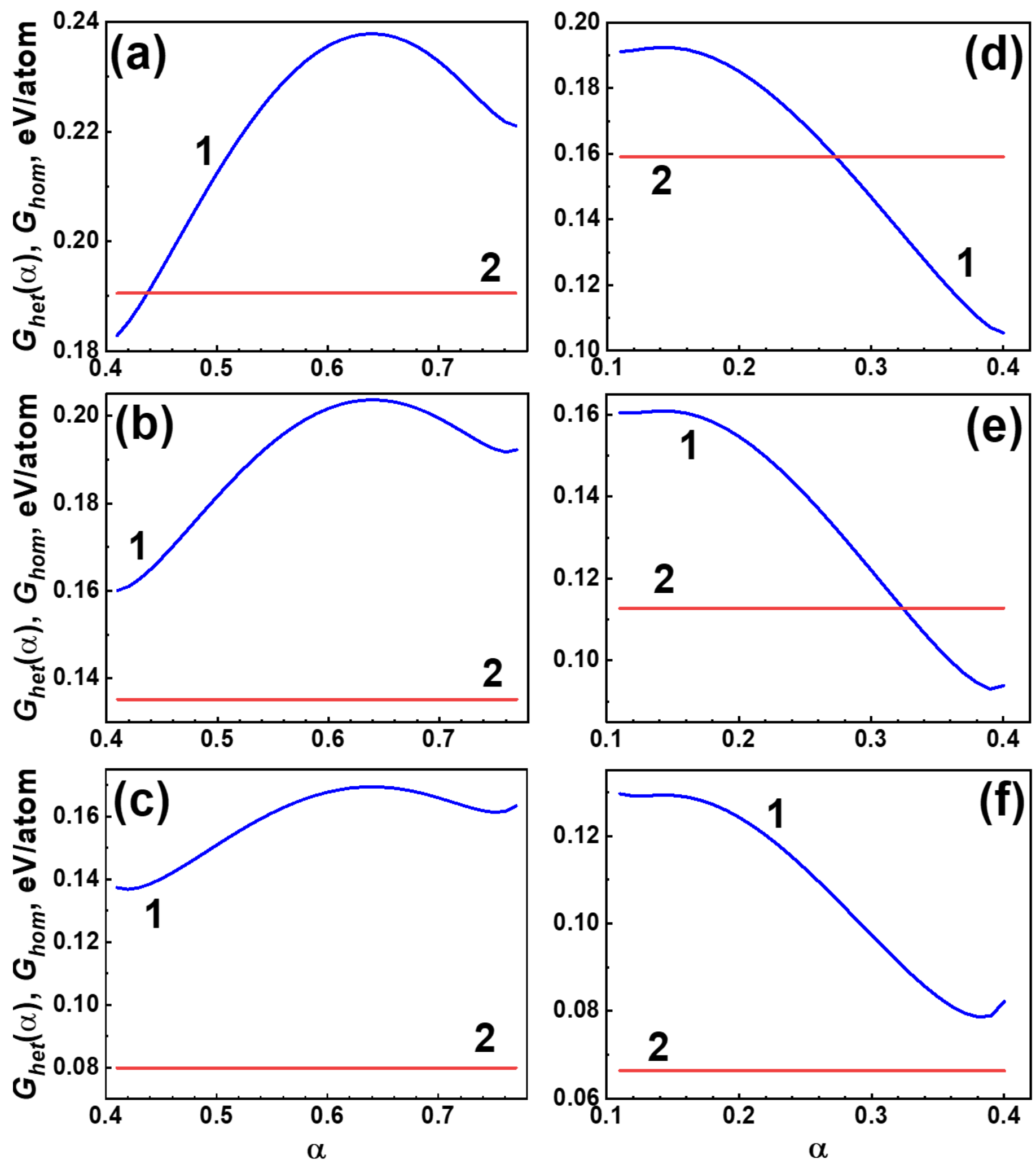

**Figure 2.** Gibbs free energy of two-phase (1) and single-phase (2) $SiO_{0.8}N_{0.3}$ (a – c) and $SiO_{0.2}N_{0.8}$ (d – f) films for the fabrication temperature *T*: (a) and (d) – 400°C, (b) and (e) – 700°C, and (c) and (f) – 1000°C.

### *3.3. Influence of $SiO_xN_y$ stoichiometry of the phase composition*

To analyze the influence of chemical composition on formation of phases in non-stoichiometric Si oxynitride films, we first set constant the relative number of oxygen atoms (stoichiometry index *x*) in the films and change the relative nitrogen content (stoichiometry index *y*). Figure 3 shows the dependences of the Gibbs free energy on the parameter α for $SiO_xN_y$ films with $x = 0.5$ and different values of *y* calculated at a temperature $T = 400$°C. It can be seen from this figure that at low *y* beginning from zero, the considered films are predicted to form two-phase states with close to stoichiometric $SiO_2$ phase (the lowest values of the Gibbs free energy at $\alpha_{min} \rightarrow \min \alpha$).

Increase in the $y$ value induces first conversion to the two-phase state with nearly stoichiometric $Si_3N_4$ and non-stoichiometric Si oxide phases (the lowest values of the Gibbs free energy at $\alpha_{min} \to \max \alpha$) at certain value of $y$ between 0.3 and 0.4. Further increase in $y$ above about 0.5 makes the films transform into a single-phase state which does not change anymore.

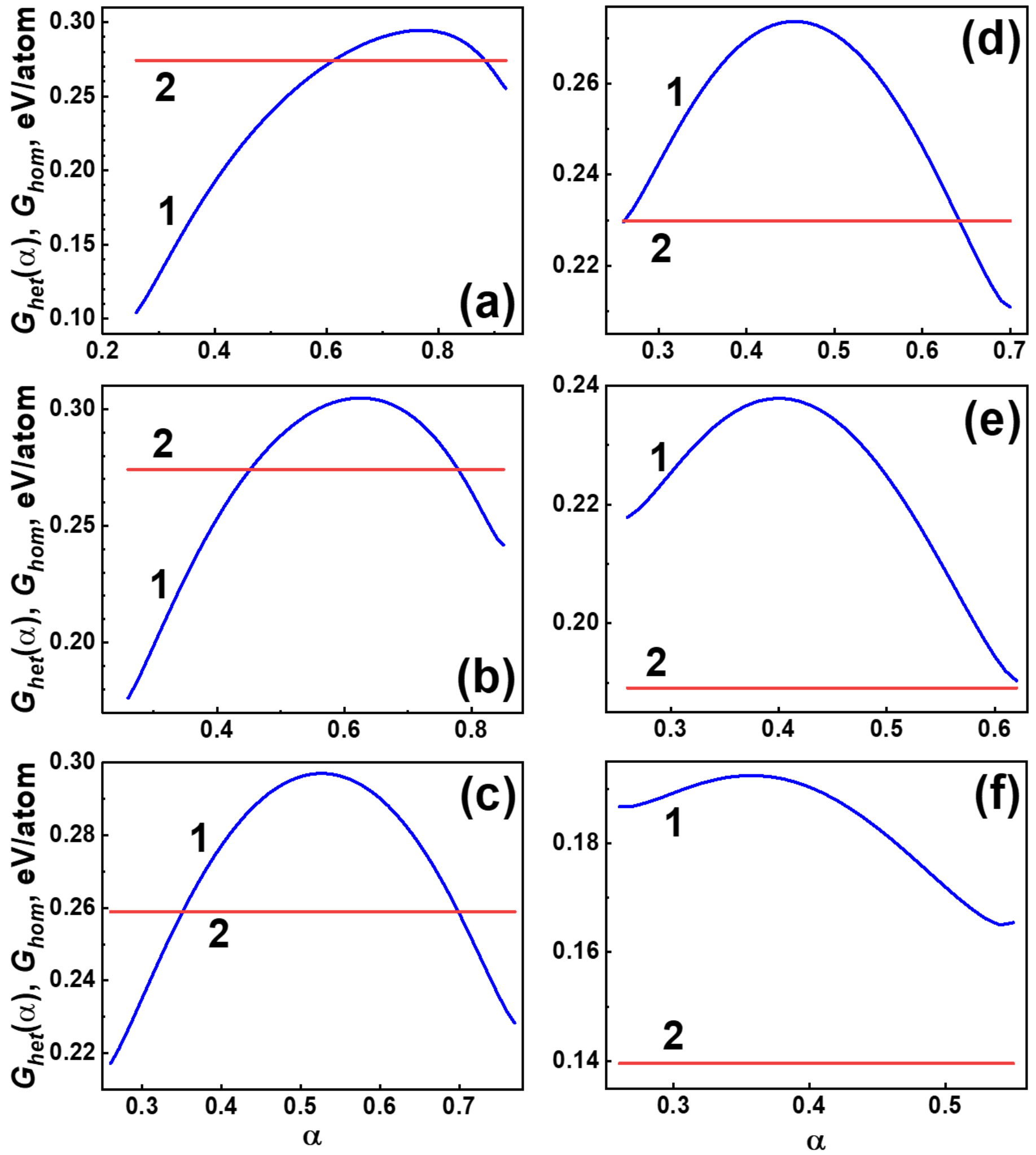

**Figure 3.** Gibbs free energy of two-phase (1) and single-phase (2) $SiO_{0.5}N_y$ films at $T = 400°C$ for $y$ values: (a) – 0.1, (b) – 0.2, (c) – 0.3, (d) – 0.4, (e) – 0.5, and (f) – 0.6.

Increase in the stoichiometry index $x$ may lead to a different behavior of the $SiO_xN_y$ phase composition at increasing the $y$ value. Figure 4 shows the dependences of the Gibbs free energy on the parameter α for the $SiO_{0.8}N_y$ films with different nitrogen contents at $T = 400°C$. It can be seen from this figure that no modification of the phase composition in the two-phase system at growing $y$ is observed in this case but a two-phase state with close to stoichiometric $SiO_2$ phase obtained at small values of $y$ transforms directly into a single-phase state when $y$ exceeds approximately 0.4.

The calculations show that the transition from the first to the second type of the transformations of phase composition of non-stoichiometric Si oxynitride films at fixed values of $x$ and increasing $y$ takes place at a certain critical value $x = x_{cr}$, which decreases at the increase in the film fabrication temperature. The reason for such dependence of $x_{cr}$ on temperature will be discussed in Section 4 with reference to the phase diagram of Si oxynitride systems.

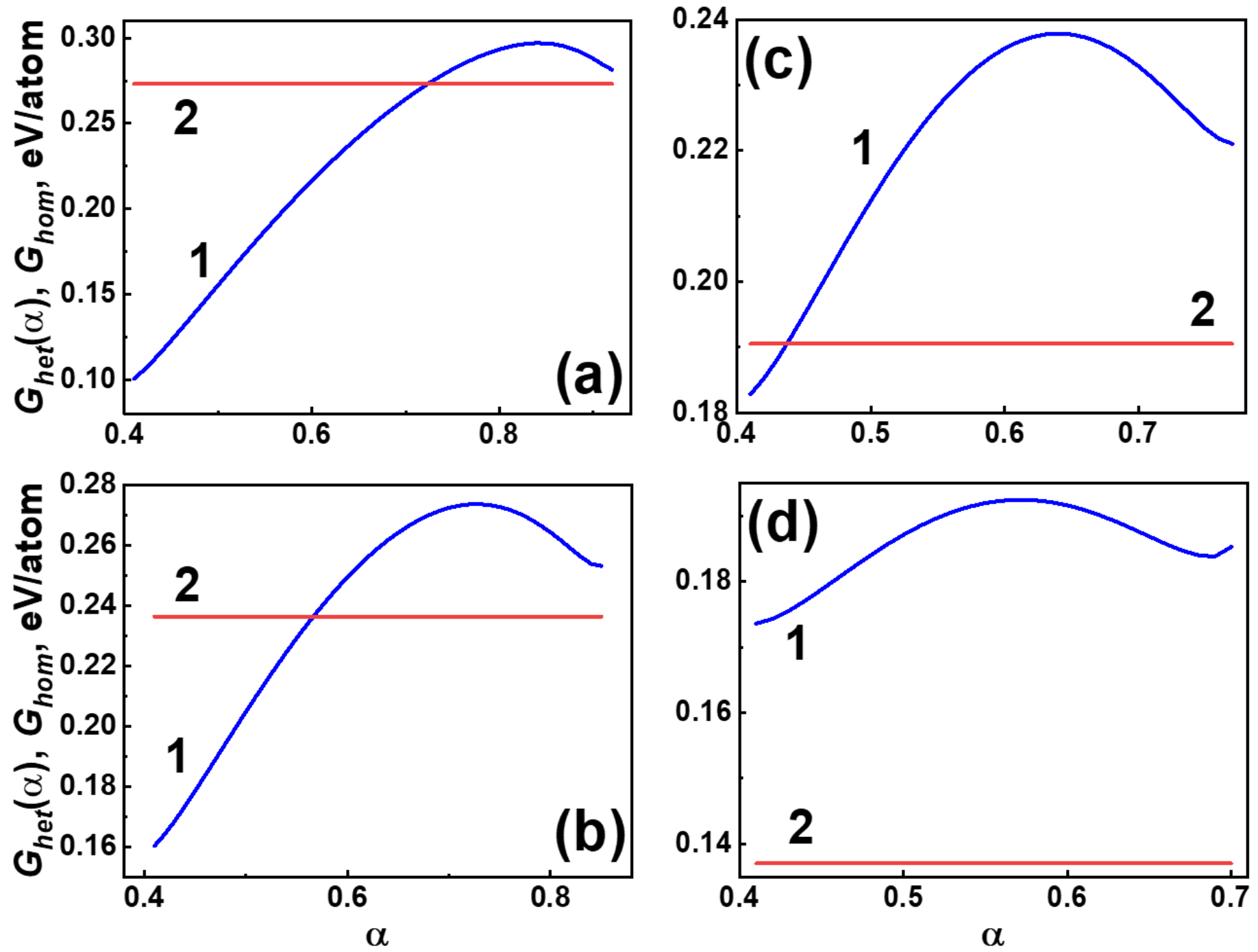

**Figure 4.** Gibbs free energy of two-phase (1) and single-phase (2) $SiO_{0.8}N_y$ films at $T = 400°C$ for $y$ values: (a) – 0.1, (b) – 0.2, (c) 0.3, and (d) – 0.4.

Similar transformations of the $SiO_xN_y$ phase composition are observed when we fix the stoichiometry index $y$ and change the value of $x$. The only difference to the previous cases is that the two-phase system formed at the lowest $x$ values consists of close to stoichiometric $Si_3N_4$ and non-stoichiometric Si oxide phases. The considered types of the evolution of $SiO_xN_y$ phase composition are illustrated by Figures 5 and 6, which present the dependences of the Gibbs free energy on the parameter α for exemplary $SiO_xN_{0.3}$ and $SiO_xN_{0.8}$ films, respectively, calculated for the fabrication temperature $T = 400°C$. As can be seen from Figure 5, at $y = 0.3$, increase in the $x$ value leads to the transition to a two-phase system with nearly stoichiometric $SiO_2$ and non-stoichiometric Si nitride at $x$ between 0.3 and 0.5 and further to a single-phase system at $x$ between 0.7 and 0.9. In its turn, Si oxynitride with $y = 0.8$ exhibits transformation of a two-phase state with close to stoichiometric $Si_3N_4$ phase directly to a single-phase state when $x$ increases to the threshold value somewhat lower than 0.3. Similar to the cases of the constant $x$ values, transition of the character of phase transformation upon increasing $x$ at a fixed value of $y$ is observed for a certain critical value $y = y_{cr}$ decreasing with an increase in the film fabrication temperature as will be discussed in Section 4.

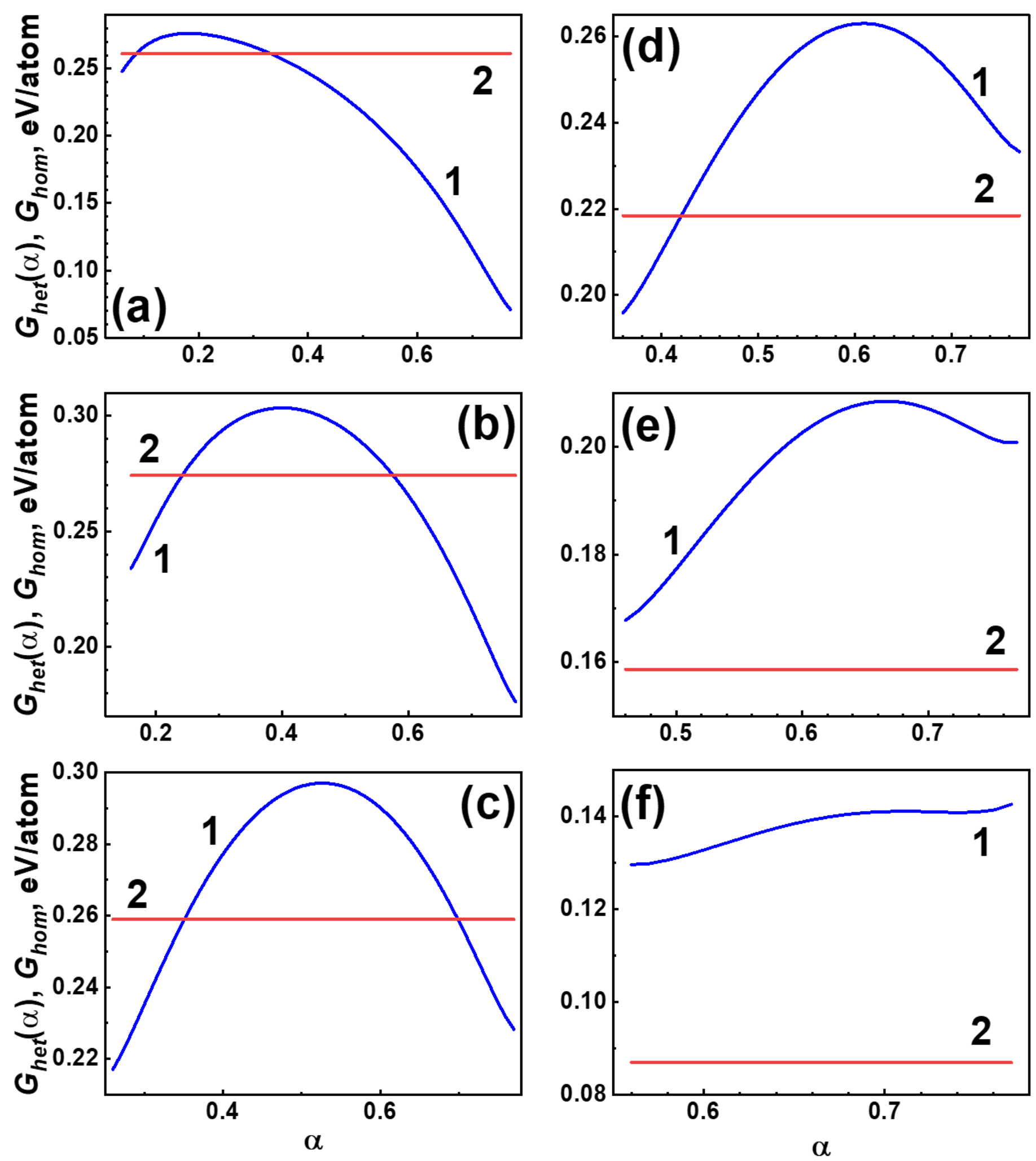

**Figure 5.** Gibbs free energy of two-phase (1) and single-phase (2) $SiO_xN_{0.3}$ films at $T = 400$°C for $x$ values: (a) – 0.1, (b) – 0.3, (c) – 0.5, (d) – 0.7, (e) – 0.9, and (f) – 1.1.

## 4. Discussion

### *4.1. Phase diagram of Si oxynitride*

The peculiarities of the formation of phase compositions of non-stoichiometric Si oxynitride films described in the previous section can be better understood based on the $SiO_xN_y$ phase diagram. We have to remark that the phase diagrams presented here correspond only to "frozen" states of the freshly fabricated films and do not take into account transformation processes that may occur after fabrication such as phase separation with formation of Si nanoprecipitates induced by high-temperature anneals [19, 52]. The influence of Si precipitation on the $SiO_xN_y$ phase composition will be considered in our forthcoming publications. Moreover, the phase diagrams cannot account for a possibility of partial Si clustering into amorphous Si inclusions since minimization of the Gibbs free energy for such cases would provide completely separated states with precipitated maximum possible excess Si concentration.

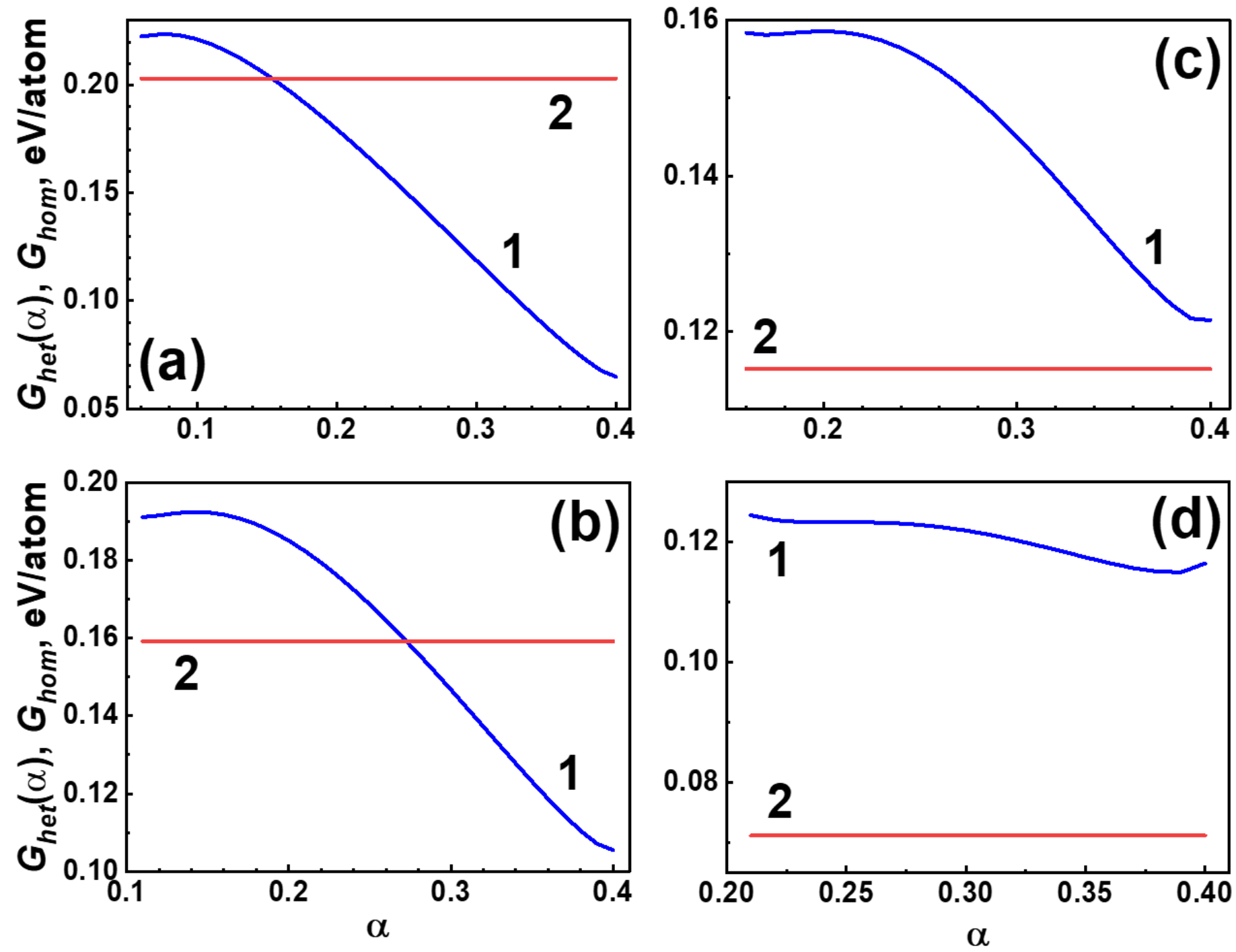

**Figure 6.** Gibbs free energy of two-phase (1) and single-phase (2) $SiO_xN_{0.8}$ films at $T$ = 400°C for $x$ values: (a) – 0.1, (b) – 0.2, (c) 0.3, and (d) – 0.4.

To obtain the phase diagram, the $SiO_xN_y$ Gibbs free energies at different stoichiometries and temperatures were analyzed. An example of the phase diagram corresponding to the fabrication temperature of 400°C is shown in Figure 7(a). In view of the nonlinearity of the expressions (1) to (7), most of the calculations were performed numerically. As can be seen from Figure 7(a), the phase diagram contains three regions corresponding to different $SiO_xN_y$ phase compositions. These regions are separated by the dependences between the critical values of $x$ and $y$ indicating the transition states (lines 2, 3 and 7). The background for calculating all the $y(x)$ dependences expressing the boundary lines is illustrated in Figures 7(b) to 7(g). From high stoichiometry indices, the phase diagram is limited by the dependence $y_{max} = \frac{4-2x}{3}$ (line 1 in Figure 7(a)) corresponding to formation of Si oxynitrides without Si excess (only Si–O and Si–N bonds). No stoichiometry indices above these values are allowed which reflects absence of excess oxygen and nitrogen in the considered $SiO_xN_y$ systems.

The lines 2 and 3 in Figure 7(a) separate the regions of the $SiO_xN_y$ stoichiometries corresponding to formation of single- and two-phase systems at a given fabrication temperature. As can be seen from Figures 7(b) and 7(c), the dependences $y(x)$ depicted by these lines are obtained as those, at which the absolute minimum of the Gibbs free energy of the two-phase system is equal to the energy of the single-phase system, $G_{hom} = \min_{\alpha} G_{het}(\alpha)$. Single-phase Si oxynitrides are predicted to form when the stoichiometry indices exceed these critical values. At smaller stoichiometries, Si oxynitrides separated into Si oxide and Si nitride phases should form. The following general trend of the $SiO_xN_y$ phase composition may be formulated: increase in the Si excess (decrease of the stoichiometry indices $x$ and $y$) favors formation of separated Si oxynitrides, while decrease in the Si excess results in homogeneous single-phase compositions, in full agreement with the conclusion made based on previous experimental and theoretical investigations [40].

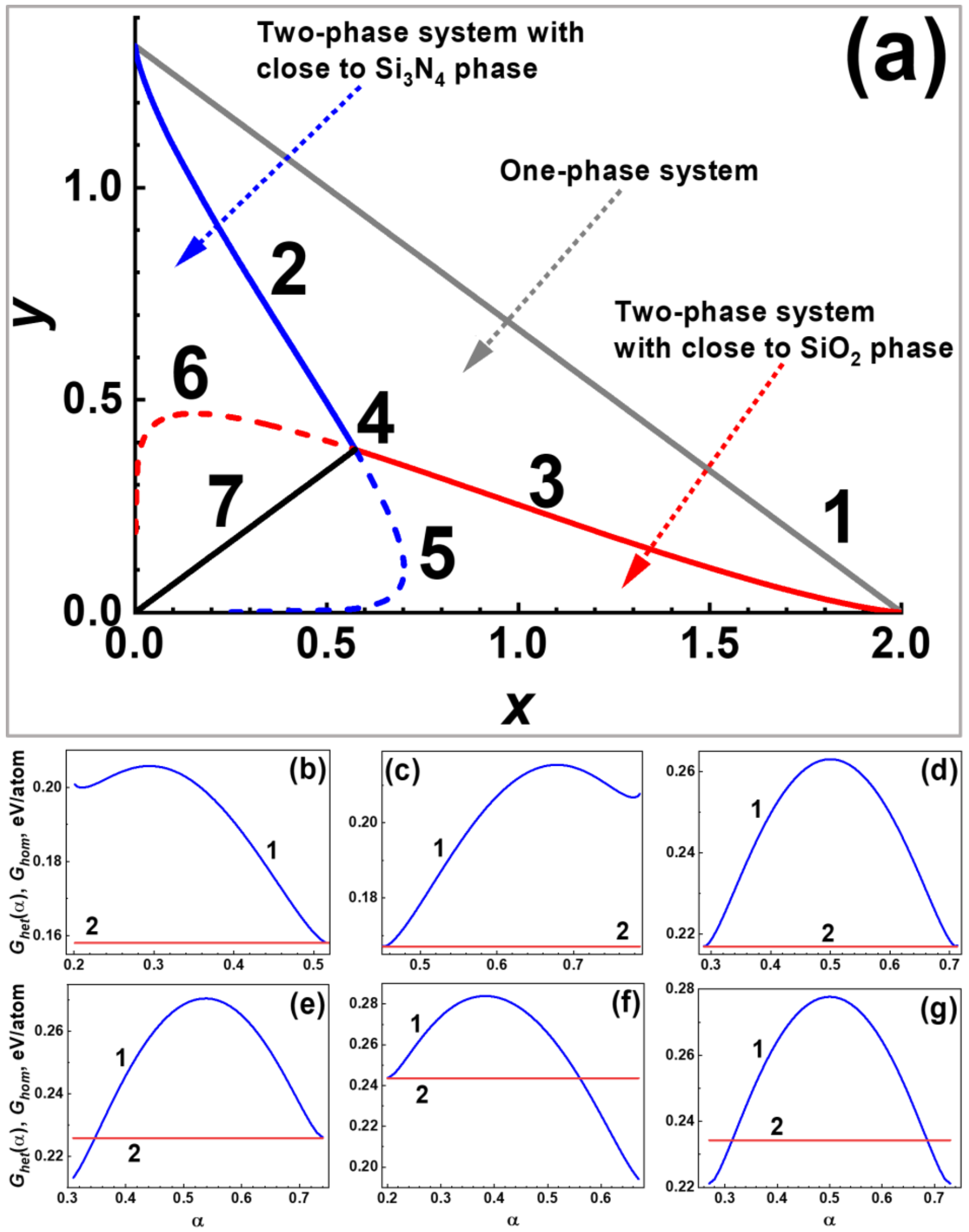

**Figure 7.** (a) – $SiO_xN_y$ phase diagram at 400°C: 1 – the highest-stoichiometries boundary of the phase diagram described by the dependence $y = \frac{4-2x}{3}$, 2 – the boundary between the regions of two-phase Si oxynitride with close to stoichiometric $Si_3N_4$ phase and single-phase Si oxynitride, 3 – the boundary between the regions of two-phase Si oxynitride with close to stoichiometric $SiO_2$ phase and single-phase Si oxynitride, 4 – the point of coexistence of two-phase Si oxynitrides with close to stoichiometric $Si_3N_4$ and close to stoichiometric $SiO_2$ phases and single-phase Si oxynitride, 5, 6 – virtual continuations of the lines 2 and 3, respectively, 7 – the boundary between the regions of two-phase Si oxynitride with close to stoichiometric $Si_3N_3$ and close to stoichiometric $SiO_2$ phases defined by the dependence $y = \frac{2}{3}x$. (b)-(g) – exemplary representations of the approaches to calculate the stoichiometries $x$ and $y$ corresponding to the lines 2 (b) and 3 (c), point 4 (d), and the lines 5 (e), 6 (f) and 7 (g) in panel (a).

As can be further seen from Figures 7(a) to 7(c), the line 2 corresponds to a boundary with the region of two-phase Si oxynitride with close to stoichiometric $Si_3N_4$ phase (right absolute minimum of the Gibbs free energy of the two-phase system). It has the origin at the point $(0, \frac{4}{3})$ corresponding to pure $Si_3N_4$. At the stoichiometry values located on the line 2, there is no preference of formation of either single-phase Si oxynitride or a two-phase one, where the Si nitride has close to $Si_3N_4$ composition. In its turn, the line 3 separates the region of two-phase Si oxynitride with close to stoichiometric $SiO_2$ phase (left absolute minimum of the Gibbs free energy of the two-phase system) from the region of single-phase Si oxynitride. Again, either homogeneous single-phase Si oxynitride or the one separated into close to $SiO_2$ and Si-rich Si nitride phases can equally form at the stoichiometry indices belonging to the line 3. The final point of this line has the coordinates (2, 0), which correspond to formation of pure $SiO_2$ phase.

The intersection point of the lines 2 and 3 (point 4 in Figure 7(a)) provides the $x$ and $y$ values, at which the Gibbs free energy of the two-phase system in the left and the right minima is the same and equal to the Gibbs free energy of the homogeneous single-phase Si oxynitride (see Figure 7(d)). For such $x$ and $y$ values, three possible states, namely single-phase Si oxynitride, and $SiO_xN_y$ separated into close to stoichiometric $SiO_2$ and Si-rich Si nitride phases or close to stoichiometric $Si_3N_4$ and Si-rich Si oxide phases can be obtained with equal probabilities.

Figures 7(e) and 7(f) show that the lines 5 and 6 in Figure 7(a) are obtained if we continue to equalize the Gibbs free energy of the single-phase Si oxynitride with the Gibbs free energy of the two-phase system in the right and left minimum, respectively, beyond the intersection point 4. It should be noted, however, that considering $SiO_xN_y$ favored by thermodynamics, we have to deal only with absolute minima of the Gibbs free energy of the two-phase systems. Therefore, the lines 5 and 6 are not a part of the phase diagram of the real Si oxynitrides and should be ignored.

The line 7 in Figure 7(a) separates the composition regions corresponding to formation of two-phase $SiO_xN_y$ with close to stoichiometric $Si_3N_4$ and close to stoichiometric $SiO_2$ phases. At the stoichiometry indices belonging to this line, the Gibbs free energy of the two-phase Si oxynitride has equal values in the left and right minima (see also Figure 7(g)). The calculations show that this equality takes place at $y = \frac{2}{3}x$. The origin of such functional dependence can be understood as follows. In the two-phase systems being considered, a Si atom has equal probabilities to contribute to either oxygen- or nitrogen-containing phase. The amount of Si forming the close to $SiO_2$ phase is almost equal to $N_{Si} = \frac{2x}{4}$, and $N_{Si} = \frac{3y}{4}$ Si atoms form the close to $Si_3N_4$ phase. Since these two states have the same formation probabilities, the numbers of Si atoms in each state should be equal, i.e. $\frac{2x}{4} = \frac{3y}{4}$, which results in $y = \frac{2}{3}x$.

The effect of temperature on the $SiO_xN_y$ phase diagram is illustrated in Figure 8. It can be seen from this figure that increase in the temperature leads to widening of the region corresponding to formation of single-phase Si oxynitrides. As only the configuration entropy contributions to the Gibbs free energy are dependent on temperature (see expressions (1) and (5)), their values for the single- and two-phase systems should be compared. The calculations show that the observed widening takes place due to faster decrease with temperature of the entropy-related contribution to the Gibbs free energy of the single-phase Si oxynitride as compared to the two-phase case. It should be also noted that since the dependence $y = \frac{2}{3}x$ (line 7 in Figure 7(a)) corresponds to equal values of the Gibbs free energy of two-phase systems in both minima, the transition point from the two-phase system with close to stoichiometric $Si_3N_4$ phase to the one with close to stoichiometric $SiO_2$ phase on the single-phase/two-phase boundary (point 4 in Figure 7(a)) is always located on this line at any $SiO_xN_y$ fabrication temperature (line 5 in Figure 8).

The limiting case $T \rightarrow \infty$ would correspond to formation of only single-phase Si oxynitrides with the boundary lines 2 and 3 (see Figure 7(a)) coinciding with the $y$ and $x$ axes, respectively. We remind again that the discussion in this work is carried out only in terms of the predictions of the thermodynamic theory described in Section 2 and does not take into account other effects that may be observed in real films such as phase separation with formation of Si nanoprecipitates induced by annealing at high temperatures, melting or even evaporation of the films etc. On the other hand, decrease of the temperature to the other limiting value, $T \rightarrow 0$ K, would make the boundaries of the two-phase Si oxynitride region turn into a single line $y = \frac{4-2x}{3}$ coinciding with the upper-stoichiometry limit of the phase diagram (line 4 in Figure 8). At this temperature, the transition point from the systems with close to $Si_3N_4$ phase to the ones with close to $SiO_2$ phase is calculated by intersecting the functions $y = \frac{2}{3}x$ and $y = \frac{4-2x}{3}$ to be at $x = 1$ and $y = \frac{2}{3}$. In this case, any non-stoichiometric $SiO_xN_y$ composition would enable formation of only two-phase Si oxynitride. For Si oxynitrides with the compositions corresponding to the limiting case $y = \frac{4-2x}{3}$, formation of either separated into fully stoichiometric $SiO_2$ and $Si_3N_4$ phases system (in view of the absence of entropy contributions to the Gibbs free energy), or homogeneous single-phase Si oxynitride would be equally possible. It should be added as well that this is the only case when two-phase stoichiometric (no Si excess) Si oxynitride may be predicted. At $T > 0$ K, only single-phase $SiO_xN_y$ with $y = \frac{4-2x}{3}$ can form. This result has a confirmation from a number of experimental studies [35, 36], where the structure of stoichiometric $SiO_xN_y$ films obeying the random bonding model has been demonstrated.

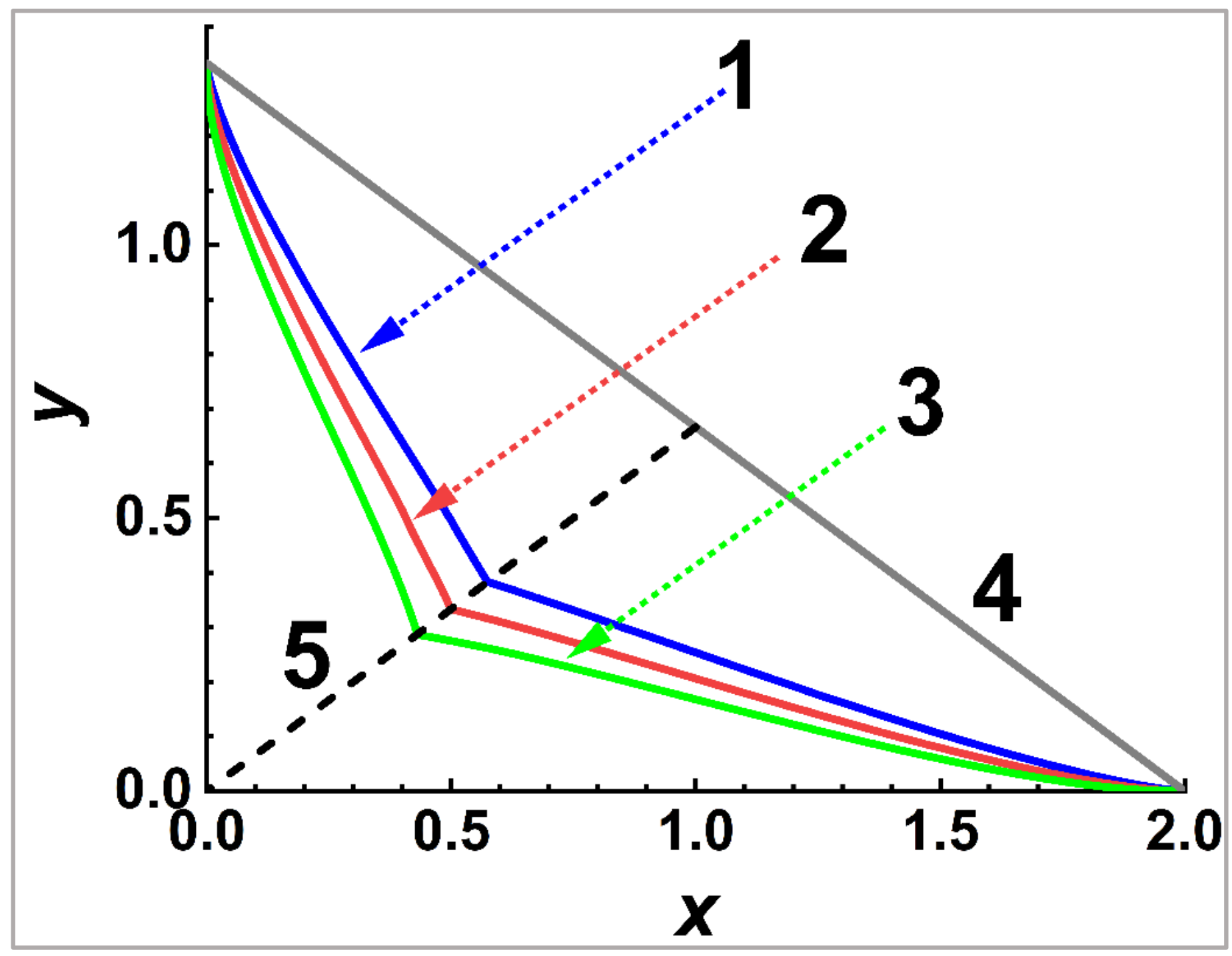

**Figure 8.** $SiO_xN_y$ phase diagrams corresponding to different fabrication temperatures: 1 – $T$ = 400°C, 2 – $T$ = 700°C, 3 – $T$ = 1000°C, and 4 – $T \rightarrow -273.15$°C (0 K). The line 4 is described by the equation $y = \frac{4-2x}{3}$ and is the highest-stoichiometries boundary of the phase diagram. The line 5 described by the equation $y = \frac{2}{3}x$ contains the points of coexistence of two-phase Si oxynitrides with close to stoichiometric $Si_3N_4$ and close to stoichiometric $SiO_2$ phases and single-phase Si oxynitride at any temperature.

### *4.2. Peculiarities of formation of phases at varying temperature and $SiO_xN_y$ composition*

In this section, the peculiarities of formation of $SiO_xN_y$ phase composition at varying the film fabrication temperature and one of the stoichiometry indices at a fixed value of the other one are analyzed.

It is demonstrated in Section 3.2 that increase in the fabrication temperature of $SiO_xN_y$ films changes the film phase composition from a two-phase to a homogeneous single-phase one (see Figure 2). At this, the transition temperature depends on the chemical composition of the films. The mechanism of this effect can be understood by analyzing the temperature dependence of the $SiO_xN_y$ phase diagram. Figure 9(a) shows the phase diagrams for three different temperatures, namely 400, 700 and 1000°C, which repeat the data presented in Figure 8. The symbols indicate the positions of the exemplary Si oxynitride compositions, namely $SiO_{0.8}N_{0.3}$ and $SiO_{0.2}N_{0.8}$, which were used to obtain the trends depicted in Figure 2. Figures 9(b) and (c) present enlarged views of the neighborhood of these compositions on the phase diagram for better clarity.

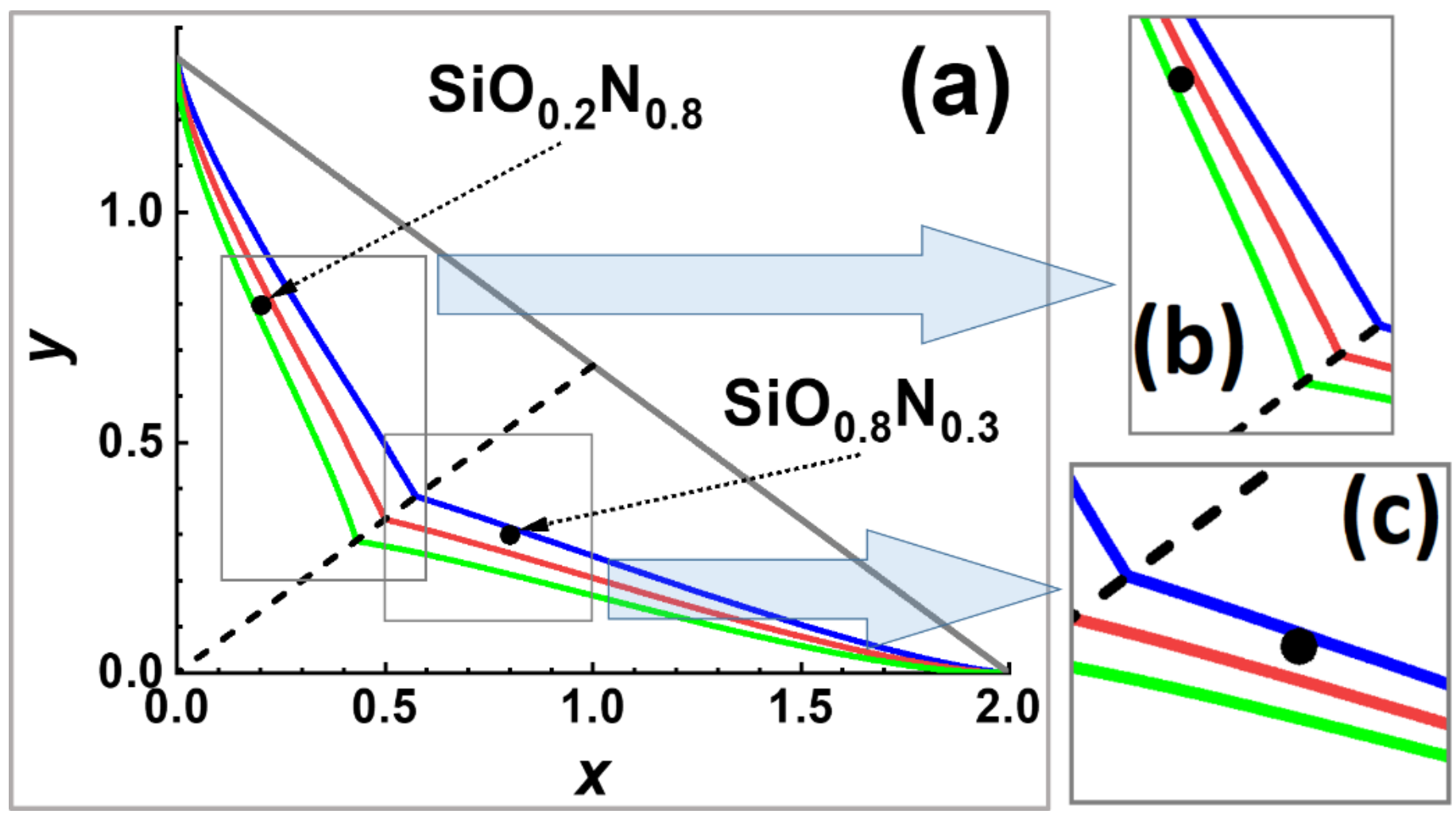

**Figure 9.** (a) – Effect of fabrication temperature on $SiO_xN_y$ phase composition illustrated for $SiO_{0.8}N_{0.3}$ and $SiO_{0.2}N_{0.8}$ as exemplary cases. (b) and (c) – enlarged neighborhoods of the points corresponding to the $SiO_{0.8}N_{0.3}$ and $SiO_{0.2}N_{0.8}$ film stoichiometries.

As can be seen from Figure 9, the transition from a two-phase Si oxynitride to a single-phase one is achieved due to expansion of the single-phase region upon temperature increase. As mentioned above, the mechanism of this expansion is the faster decrease of the entropy contribution to the Gibbs free energy of single-phase Si oxynitride as compared to that for two-phase systems. This effect equally refers to the two-phase systems with close to stoichiometric $SiO_2$ and close to stoichiometric $Si_3N_4$ phases. Given the presented examples, the symbols representing $SiO_{0.8}N_{0.3}$ and $SiO_{0.2}N_{0.8}$ are located left-lower from the boundary between the two-phase and single-phase regions for the temperature of 400°C, which corresponds to formation of two-phase Si oxynitrides in both cases (see also Figure 2(a) and (d)). At 700°C, the $SiO_{0.8}N_{0.3}$ point becomes right-higher from the respective two phase/single phase boundary, indicating formation of already a single-phase system, in compliance with the result presented in Figure 2(b). On the other hand, the symbol marking the $SiO_{0.2}N_{0.8}$ composition remains left-lower from the two phase/single phase boundary. Hence, Si oxynitride with such stoichiometry is still a two-phase system confirming the result of Figure 2(e). Finally, at 1000°C, the symbols for both considered stoichiometries become right-higher than the two phase/single phase boundary meaning formation of homogeneous single-phase systems (see also Figure 2(c) and (f)). As can be further seen from Figure 9, no change in the phase composition within the two-phase region (no transitions between Si oxynitrides with close to $SiO_2$ and close to $Si_3N_4$ phases) may take place at varying the $SiO_xN_y$ fabrication temperature, as the phase composition of the two-phase system is defined solely by the position of the point ($x$, $y$) with respect to the line $y = \frac{2}{3}x$, which has no dependence on temperature.

The effects of variation of one of the stoichiometry indices at a fixed value of the other index on the $SiO_xN_y$ phase composition enabling to better understand the results shown in Figures 3-6 are illustrated in Figure 10. It can be seen from Figure 10(a) that transformation of the phase composition at a fixed value of the stoichiometry index $x$ and increasing the $y$ value depicted in Figure 3 is obtained at $x$ smaller than the critical value corresponding to the point of coexistence of both kinds of two-phase systems and a single-phase system (point 4 in Figure 7(a)). In this case, raising the $y$ value from the smallest ones shifts the point representing the $SiO_xN_y$ composition on the phase diagram from the region of two-phase Si oxynitrides with close to stoichiometric $SiO_2$ first to the region of two-phase systems with close to stoichiometric $Si_3N_4$ (the transition 1→2 in Figure 10(a)) and then to the region corresponding to formation of single-phase Si oxynitrides as illustrated by the transition 2→3 in Figure 10(a). As can be further seen from Figure 10(a), at $x$ larger than the above-mentioned critical value corresponding to coexistence of three possible $SiO_xN_y$ phase states, only transitions from a two-phase Si oxynitride with close to stoichiometric $SiO_2$ directly to a single-phase state are possible, as illustrated by the transition 4→5. It should be also noted that the type of transition from two- to single-phase state is temperature-dependent. Namely, since the critical $x$ value decreases with the increase in the $SiO_xN_y$ fabrication temperature (see Figure 8), raise of the temperature modifies the phase states transition from the first to the second type.

Analogous situation is obtained when we fix the $y$ value in the $SiO_xN_y$ composition and increase $x$ from the smallest values. This case is illustrated by Figure 10(b) for a number of stoichiometry indices used to obtain the dependences presented in Figures 5 and 6 as exemplary cases. At this, the smallest values of $x$ always correspond to two-phase Si oxynitride states with close to stoichiometric $Si_3N_4$ phase. Transformation into a single-phase system with an intermediary two-phase state with close to stoichiometric $SiO_2$ upon raising the $x$ value takes place when the $y$ value is smaller than the critical one for coexistence of three possible $SiO_xN_y$ states (the transformation 1→2→3 in Figure 10(b)). Otherwise, a two-phase Si oxynitride with close to stoichiometric $Si_3N_4$ turns directly into a single-phase system as demonstrated by the transformation 4→5 in Figure 10(b). Again, raising the fabrication temperature of the $SiO_xN_y$ films induces change from the first to the second type of the transition due to the decrease of the critical value of $y$.

### *4.3. Phase compositions of PECVD grown $SiO_xN_y$ films*

In this section, we make a comparative analysis of the phase compositions of a number of $SiO_xN_y$ films with different stoichiometries obtained from the calculated phase diagram and inferred based on interpretation of the film FTIR absorption spectra [40, 47]. Analysis of the FTIR spectra provides insights into the distribution of phases in the $SiO_xN_y$ films depending on their chemical composition [53]. At this, we fully realize that the conclusions based on the results of the FTIR spectroscopy alone cannot be absolutely convincing. Complex investigations involving other methods, in particular XPS, in addition to FTIR spectroscopy will enable a more comprehensive characterization of the phase composition of the investigated Si oxynitrides, and this work is planned for the future. The aim of the present analysis is to establish consistency of the phase compositions of the $SiO_xN_y$ films obtained by interpreting the FTIR spectra and the ones predicted based on the phase diagram to support the theoretical model of formation of phases in Si-rich Si oxynitrides.

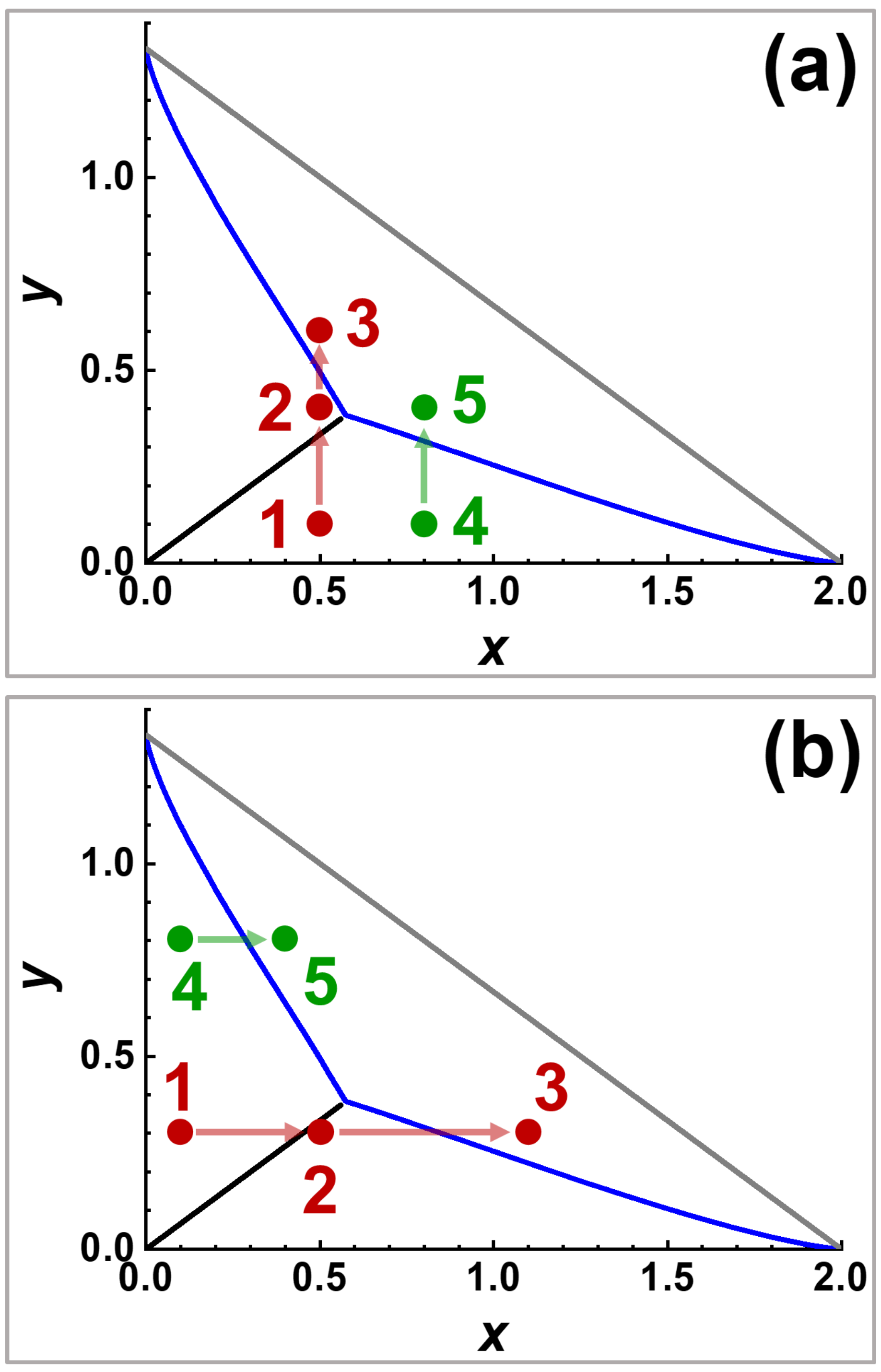

**Figure 10.** (a) – Effects of varying the stoichiometry index *y* at a fixed value of *x* on the $SiO_xN_y$ phase composition. The $SiO_xN_y$ stoichiometries: 1 – $SiO_{0.5}N_{0.1}$, 2 – $SiO_{0.5}N_{0.4}$, 3 – $SiO_{0.5}N_{0.6}$, 4 – $SiO_{0.8}N_{0.1}$, and 5 – $SiO_{0.8}N_{0.4}$. (b) – Effects of varying the stoichiometry index *x* at a fixed value of *y* on the $SiO_xN_y$ phase composition. The $SiO_xN_y$ stoichiometries: 1 – $SiO_{0.1}N_{0.3}$, 2 – $SiO_{0.5}N_{0.3}$, 3 – $SiO_{1.1}N_{0.3}$, 4 – $SiO_{0.1}N_{0.8}$, and 5 – $SiO_{0.5}N_{0.8}$.

The investigated $SiO_xN_y$ films were PECVD-grown at 320°C on double-side polished boron-doped Czochralski Si (100) wafers. The film stoichiometries were determined by time-of-flight secondary ion mass spectrometry (SIMS). The FTIR measurements were performed in a transmission mode at normal incidence with a Si substrate as a reference sample using a PerkinElmer BX-II spectrometer. The measurement resolution was 2 $cm^{-1}$. The IR transmission signals were converted into the absorption signals using the Beer-Bouguer-Lambert law. The absorption spectra in the wavenumber range of 600 to 1300 $cm^{-1}$ were mathematically deconvoluted into elementary Gaussian profiles corresponding to stretching vibrations of Si–O and Si–N bonds [40], taking into account the Alentsev-Fock criterion [54] as well as using our experience in analyzing the structure and composition of non-stoichiometric Si oxides [55, 56]. Full details concerning film preparation and characterization can be found in [40].

Figure 11(a) shows the calculated phase diagram corresponding to the $SiO_xN_y$ fabrication temperature of 320°C. The symbols 1 to 3 indicate positions of three stoichiometries of the $SiO_xN_y$ films chosen for analysis, namely $SiO_{1.09}N_{0.32}$, $SiO_{0.59}N_{0.16}$ and $SiO_{0.11}N_{1.07}$, with predicted single-phase composition, and two-phase compositions with close to stoichiometric $SiO_2$ and close to stoichiometric $Si_3N_4$, respectively. It should be noted that the latter stoichiometry was not obtained by SIMS but estimated from the maximum position of the FTIR absorption spectrum and the ratio of the integrated intensities of the Si–N and Si–O related bands comparing them to the spectra of other films in the same technological series. Moreover, to additionally confirm the validity of our theoretical predictions, we have added the symbols 4 and 5 in Figure 11(a), which indicate the stoichiometries $SiO_{1.3}N_{0.28}$ and $SiO_{0.37}N_{0.11}$ of the films, whose phase compositions analyzed in [47] demonstrated good agreement with the calculated phase diagram.

Figure 11(b) to (d) shows FTIR absorption spectra of the $SiO_{1.09}N_{0.32}$, $SiO_{0.59}N_{0.16}$ and $SiO_{0.11}N_{1.07}$ samples with separated components corresponding to Si–O (blue) and Si–N (green) bond vibrations. As can be seen from this figure, the spectra differ by their maximum positions and shape, which may be interpreted in terms of different phase compositions of the respective films.

Analysis of the blue integrated band of the FTIR spectra presented in Figure 11(b) to (d) reveals that this band may be characterized by elementary Gaussian profiles corresponding to tetrahedral structural units $Si–O_aSi_{4-a}$ ($1 \leq a \leq 4$) of non-stoichiometric Si oxide or to 4- and 6-member rings of Si–$O_4$ tetrahedra of $SiO_2$ phase [57]. The green integrated bands caused by presence of Si–N bonds consist of elementary profiles with the characteristics determined in [40]. The integrated intensity of each of these bands has a direct correlation with the concentrations of the Si–O and Si–N bonds, respectively.

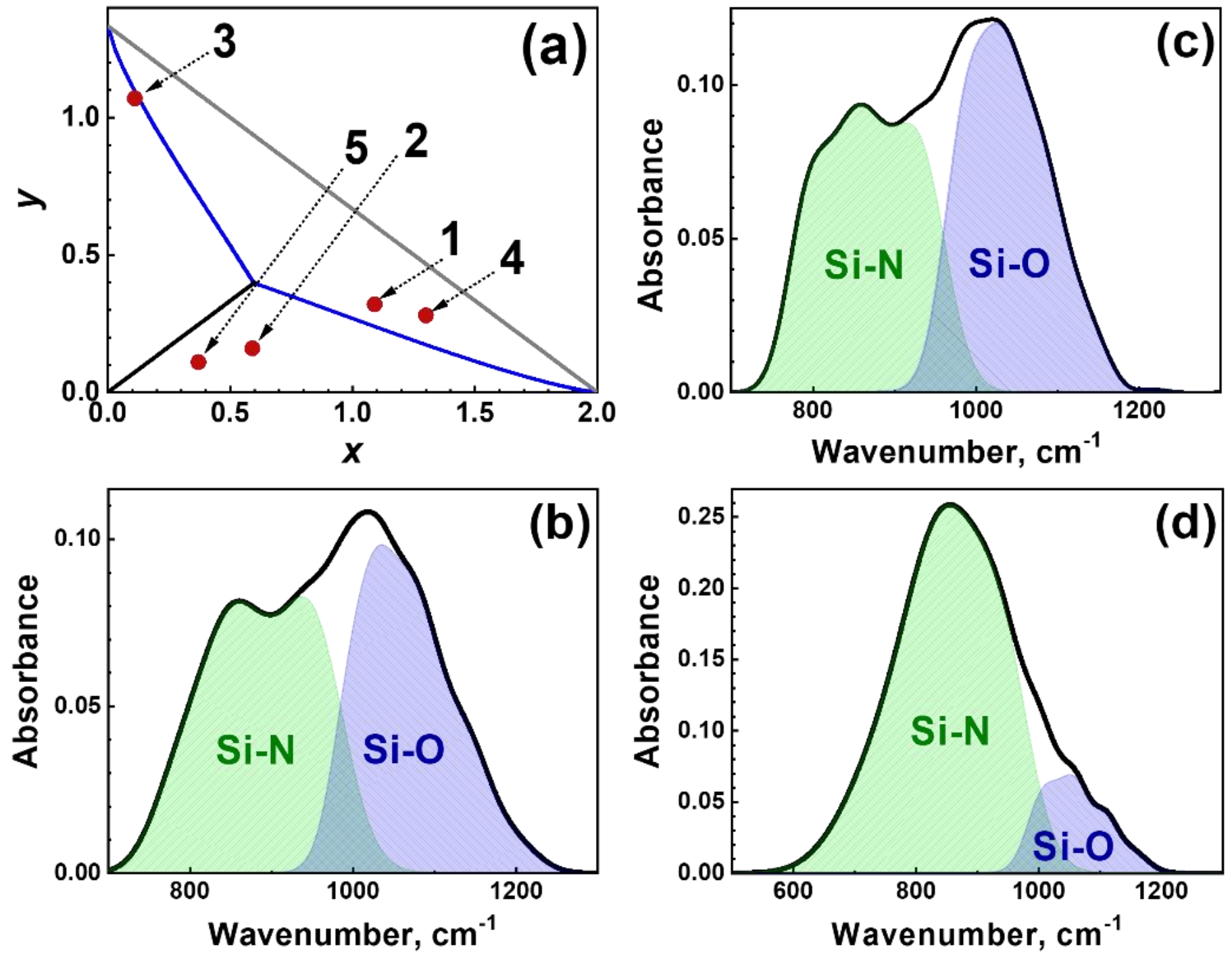

**Figure 11.** (a) – Calculated phase diagram at 320°C. The symbols correspond to the following $SiO_xN_y$ film stoichiometries: 1 – $SiO_{1.09}N_{0.32}$, 2 – $SiO_{0.59}N_{0.16}$, 3 – $SiO_{0.11}N_{1.07}$, 4 – $SiO_{1.3}N_{0.28}$, and 5 – $SiO_{0.37}N_{0.11}$. (b)-(d) – FTIR absorption spectra of the $SiO_{1.09}N_{0.32}$ (b), $SiO_{0.59}N_{0.16}$ (c) and $SiO_{0.11}N_{1.07}$ (d) films.

It can be seen from Figure 11(b) that both the Si–O and Si–N related IR bands of the $SiO_{1.09}N_{0.32}$ film are of irregular shapes and strongly overlap ensuring rather smooth transition between them. The shape of the Si–O band indicates formation of Si–$O_aSi_{4-a}$ complexes with multiple values of $a$. The strongly overlapping Si–O and Si–N related bands point to existence of mixed tetrahedral units Si–$O_aN_bSi_{4-a-b}$ that contain both oxygen and nitrogen atoms [39]. Moreover, the Si–N related band is composite thus implying distribution of the local stoichiometry of the Si–N containing complexes. These results may be interpreted as a single-phase $SiO_{1.09}N_{0.32}$ composition made up by Si–$O_aN_bSi_{4-a-b}$ ($0 \le a + b \le 4$) structural units that may simultaneously contain both O and N atoms.

On the other hand, the Si–O and Si–N related bands in the FTIR spectrum of the $SiO_{0.59}N_{0.16}$ film presented in Figure 11(c) have a higher degree of separation from each other as compared to the previous case. Moreover, the Si–O band has a symmetrical shape close to Gaussian one, which points to dominance of a single elementary component corresponding to Si–$O_aSi_{4-a}$ tetrahedral units with a certain value of *a*. Analysis of the composition of this band shows that the dominating elementary component lies within the range 1020-1060 см$^{-3}$, which corresponds to almost fully oxidized Si–$O_3$Si complexes. Prevalence of such complexes in the Si oxide structure is reasonably explained by that they form extended boundaries of small clusters with the close to stoichiometric $SiO_2$ chemical composition [58, 59]. In its turn, the shape of the Si–N related band in Figure 11(c) remains composite, which again points to the distribution of the oxidation state of central Si atoms in the tetrahedral units containing nitrogen atoms. At this, decreased overlap of the Si–O and Si–N related bands indicates smaller probability of an oxygen atom to occur in a vicinity of a nitrogen atom, i.e. formation of mixed complexes Si–$O_aN_bSi_{4-a-b}$, compared to case depicted in Figure 11(b). Therefore, the FTIR results can be reasonably explained by formation of a two-phase system with the high-stoichiometry Si oxide phase and the non-stoichiometric Si nitride phase having distribution of the oxidation degree of tetrahedral structural units, in consistency with the phase diagram presented in Figure 11(a).

The FTIR spectrum of the $SiO_{0.11}N_{1.07}$ film shown in Figure 11(d) exhibits the Si–N related band with nearly symmetrical Gaussian shape and maximum position at 855±5 cm$^{-1}$, which points to dominance of structural units with a single local stoichiometry. It is known that the IR band maximum for the films that contain Si–$N_4$ complexes is positioned at 830-850 cm$^{-3}$ [60, 61]. Increase in the oxygen concentration in the films shifts this maximum toward higher energies. One may therefore interpret the Si–N related band as originating from the Si–$N_4$ complexes of the close to stoichiometric $Si_3N_4$ phase. In its turn, the Si–O related band has irregular shape pointing to distribution of the oxidation degrees of Si–$O_aSi_{4-a}$ tetrahedral units. These results allow concluding that the $SiO_{0.11}N_{0.16}$ film is two-phase consisting of non-stoichiometric Si oxide phase and nitrogen enriched Si nitride phase with prevalence of Si–$N_4$ complexes.

It should be noted that, unlike the theoretical phase diagram that does not take into account intermediary local compositions at the boundaries of the clusters of different phases as well as deviations from ideality generally observed in real systems, the FTIR spectra are sensitive to all these effects. Moreover, as discussed above, FTIR spectroscopy alone is not capable of providing undoubtful information about the phases in the investigated $SiO_xN_y$ films. Viewed this way, the phase compositions of the $SiO_xN_y$ films obtained from the calculated phase diagram are well supported by the characteristics of the Si–O and Si–N related bands in the FTIR absorption spectra, which confirms the validity of the thermodynamic model of formation of phases in Si-rich Si oxynitrides.

We also made a comparison of the phase compositions of a number of PECVD-grown Si oxynitride films with stoichiometries in a wide range reported in literature with predictions of the thermodynamic model considered in this paper. The experimental phase compositions were determined for the films grown at 320°C [30, 41], 600°C [11], and not explicitly indicated low temperature [31]. In Figure 12(a) and (b), we plotted the calculated phase diagrams corresponding to the temperatures of 320°C and 600°C, respectively, along with the stoichiometry indices of the $SiO_xN_y$ films, with the data from [31] placed in Figure 12(a). The theoretical phase compositions and experimental structural models of the investigated films are also summarized in Table 1.

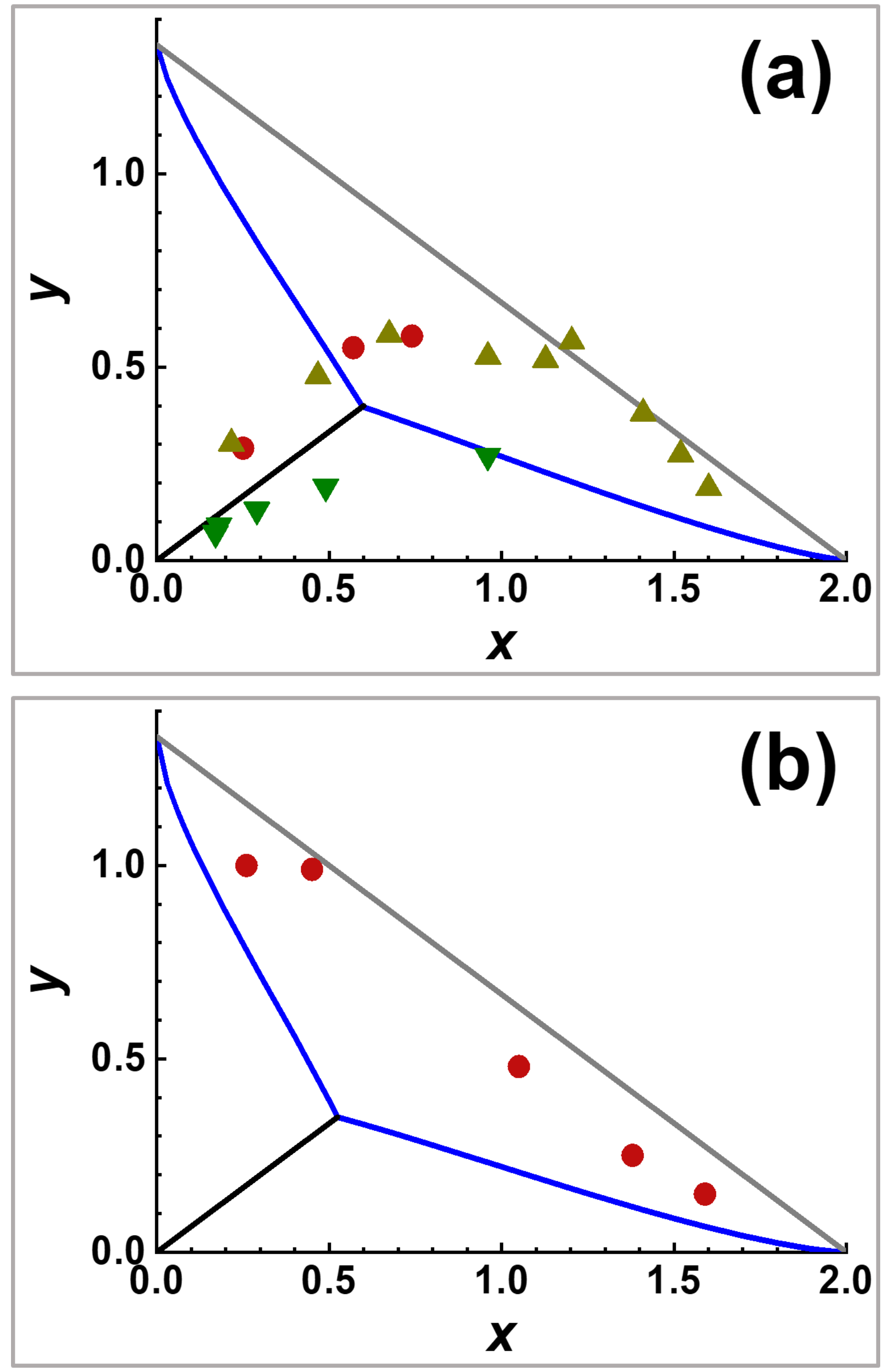

**Figure 12.** Calculated phase diagrams at 320°C (a) and 600°C (b). Symbols indicate $SiO_xN_y$ compositions from: (a) – [30] (circles), [41] (triangles up), and [31] (triangles down); and (b) – [11] (circles).

Table 1. Comparison of theoretical phase compositions and experimental structural models of PECVD-grown $SiO_xN_y$ films

| $SiO_xN_y$ composition | Theoretical phase composition | Experimental structural model | Reference |
|---|---|---|---|
| $SiO_{0.25}N_{0.29}$ | Two-phase with close to $Si_3N_4$ | RMM with phases formed by Si–Si bonds and Si–O and Si–N bonds | [30] |
| $SiO_{0.57}N_{0.55}$ | Single-phase | RMM with phases formed by Si–Si bonds and Si–O and Si–N bonds | [30] |
| $SiO_{0.74}N_{0.58}$ | Single-phase | RBM | [30] |
| $SiO_{0.22}N_{0.3}$ | Two-phase with close to $Si_3N_4$ | RBM | [41] |
| $SiO_{0.47}N_{0.48}$ | Two-phase with close to $Si_3N_4$ | RBM | [41] |
| $SiO_{0.67}N_{0.58}$ | Single-phase | RBM | [41] |
| $SiO_{0.96}N_{0.52}$ | Single-phase | RBM | [41] |
| $SiO_{1.13}N_{0.52}$ | Single-phase | RBM | [41] |
| $SiO_{1.2}N_{0.57}$ | Out of phase diagram | RBM | [41] |
| $SiO_{1.41}N_{0.38}$ | Single-phase | RBM | [41] |
| $SiO_{1.52}N_{0.27}$ | Single-phase | RBM | [41] |
| $SiO_{1.6}N_{0.19}$ | Single-phase | RBM | [41] |
| $SiO_{0.17}N_{0.07}$ | Two-phase with close to $SiO_2$ / two-phase with close to $Si_3N_4$ | RMM with close to $Si_3N_4$ phase | [31] |
| $SiO_{0.18}N_{0.09}$ | Two-phase with close to $SiO_2$ / two-phase with close to $Si_3N_4$ | RMM with close to $Si_3N_4$ phase | [31] |
| $SiO_{0.29}N_{0.13}$ | Two-phase with close to $SiO_2$ | RMM with close to $Si_3N_4$ phase | [31] |
| $SiO_{0.49}N_{0.19}$ | Two-phase with close to $SiO_2$ | Mixture of Si nitride and Si oxynitride phases | [31] |
| $SiO_{0.96}N_{0.27}$ | Two-phase with close to $SiO_2$ / single-phase | Mixture of Si oxide, sub-oxide, and oxynitride phases | [31] |
| $SiO_{0.26}N_{1.0}$ | Single-phase | RBM | [11] |
| $SiO_{0.45}N_{0.99}$ | Single-phase | RBM | [11] |
| $SiO_{1.05}N_{0.48}$ | Single-phase | RBM | [11] |
| $SiO_{1.38}N_{0.25}$ | Single-phase | RBM | [11] |
| $SiO_{1.59}N_{0.15}$ | Single-phase | RBM | [11] |

According to the results of XANES and TEM investigations [30], the $SiO_{0.74}N_{0.58}$ films have a homogeneous structure better represented by the random bonding model. Furthermore, this model also suits for interpreting the microstructure of the $SiO_xN_y$ films with $x$ ranged from 0.67 to 1.6 and $y$ ranged from 0.58 down to 0.19 [41]. The first coordination shell in such films always accommodate O and N atoms as revealed by combined RBS and EXAFS spectroscopy. The obtained experimental results fully agree with the homogeneous single-phase Si oxynitride structures with the mentioned chemical compositions predicted by the thermodynamic model (see Figure 12(a)).

Raise of the Si content induces Si clustering during $SiO_xN_y$ film formation [31-33]. Si–Si and Si–H bonds become detectable in the samples with the relative Si content over 50% [41]. Hence, the experimental structures of the $SiO_{0.57}N_{0.55}$, $SiO_{0.47}N_{0.48}$, $SiO_{0.25}N_{0.29}$, and $SiO_{0.22}N_{0.3}$ films correspond to a mixture model with the phases formed by Si–Si bonds as well as Si–O and Si–N bonds. The former two compositions in the vicinity of the two-phase/single-phase boundary are consistent with the theoretically predicted homogeneous Si oxynitride films with distributed amorphous Si clusters (see Figure 12(a)). In their turn, intermixed Si–O and Si–N bonds in the structure of the $SiO_{0.25}N_{0.29}$, and $SiO_{0.22}N_{0.3}$ films do not exactly fit to the model but may appear from extended boundaries between small-sizes Si oxide and Si nitride agglomerates.

In [31], XPS has been used to probe the chemical composition and states of the constituents of the $SiO_xN_y$ films with $0.17 \leq x \leq 0.96$ and $0.07 \leq y \leq 0.27$ grown at low temperature. Since no temperature value was provided, we compared the reported phase compositions with theoretical ones for 320°C, typical for low-temperature $SiO_xN_y$ fabrication. In all the cases, the mixture model was applied to describe the structure of the investigated films. The $SiO_{0.96}N_{0.27}$ film was represented by a complicated mixture of Si oxide, sub-oxide, and oxynitride phases, which agrees with the position of the respective stoichiometry on the phase diagram very close to the boundary between the two-phase state with close to stoichiometric $SiO_2$ and homogeneous single-phase state. The $SiO_{0.49}N_{0.19}$ was experimentally determined to be a mixture of Si oxynitride and close to stoichiometric Si nitride phases. This experimental phase composition does not fully correspond to the theoretical one. The favor of formation of nearly stoichiometric Si nitride in such films may be explained by the relative proximity of their composition to the boundary between the two-phase systems with close to $SiO_2$ and close to $Si_3N_4$ phase (see Figure 12(a)). Even higher proximity to this boundary explains the results for the rest of the films studied in this work ($SiO_{0.17}N_{0.07}$, $SiO_{0.18}N_{0.09}$ and $SiO_{0.29}N_{0.13}$), which contained nitrogen only in the close to stoichiometric Si nitride phase.

In [11], FTIR spectroscopy was used to obtain information on the short-range order for a number of $SiO_xN_y$ compositions. A smooth transition of the FTIR spectra was observed when the composition of the thin amorphous $SiO_xN_y$ layers varied from $SiO_{1.9}$ to $SiN_{1.06}$. Presence of a single main absorbance band in the spectra is indicative of the bonding configuration better described by the random bonding model. Such description fully agrees with the theoretically predicted single-phase composition as illustrated by the phase diagram in Figure 12(b).

Comparison of experimental and theoretical phase compositions of the $SiO_xN_y$ films allows one to conclude their generally good agreement. The main discrepancies between the theory and experiment appear for the film compositions with high relative Si contents, where two-phase states are expected. Analysis of the phase compositions of such films in the framework of the thermodynamic model is complicated by the presence of high amounts of Si-Si bonds belonging to amorphous Si clusters not accounted for by the model. Moreover, partial separation of the excess Si increases the stoichiometry indices of the rest Si oxynitride, which may affect interpretation of its phase composition. Therefore, the theoretical predictions may deviate from the experimentally determined phase compositions ever more with growth of the relative Si content. Furthermore, the boundaries between different phases in two-phase systems introducing both O- and N-containing structural units are not covered by the theory. Also, the reliability of predictions of the theoretical model can hardly be expected at the smallest O and N contents when the film composition should be considered no more as Si oxynitride but rather as Si with small amounts of O and N impurities.

## 5. Conclusions

In conclusion, we developed a comprehensive thermodynamic description of the formation of phase compositions during fabrication of Si oxynitride ($SiO_xN_y$, $0 \leq x \leq 2$ and $0 \leq y \leq 4/3$) films as a function of the film chemical composition and fabrication temperature. Two possible phase states of the Si oxynitride films, namely homogeneous single-phase one and two-phase represented by a mixture of Si oxide and Si nitride phases, are considered. The Gibbs free energies of $SiO_xN_y$ in both states are compared. The stability of either phase composition at given values of $x$ and $y$ as well as the temperature $T$ is inferred based on the minimum-free-energy criterion.

The study of the two-phase Si oxynitrides demonstrates one of its phases (either Si oxide or Si nitride) having chemical composition close to stoichiometric one, with the rest of the Si atoms contributing to the other (generally non-stoichiometric) phase. Such finding highlights a possibility of creating composites with Si nanoparticles embedded in different local dielectric environments by phase separation during high-temperature annealing [23, 50].

The peculiarities of transitions of the phase compositions with changes in the $SiO_xN_y$ stoichiometry indices and fabrication temperature are determined. Two-phase Si oxynitrides are favored to form at low values of the stoichiometry indices $x$ and $y$ and fabrication temperatures. Raising the temperature induces transition from a two- to a single-phase state due to the faster decrease with temperature of the entropy-related contribution to the Gibbs free energy for the single-phase $SiO_xN_y$ system compared to that for the two-phase system. Increasing one of the stoichiometry indices ($x$ or $y$) at the fixed value of the other one also favors $SiO_xN_y$ transition from a two-phase to a single-phase state. This transition may be direct or involve an intermediary change of the phase composition of the two-phase system (such as close to stoichiometric $SiO_2$ + non-stoichiometric Si nitride → close to stoichiometric $Si_3N_4$ + non-stoichiometric Si oxide or vice versa) depending on whether the value of the fixed stoichiometry index exceeds or is below the critical value. A comprehensive phase diagram indicating the stoichiometry regions corresponding to single-phase, two-phase with close to stoichiometric $SiO_2$ and two-phase with close to stoichiometric $Si_3N_4$ Si oxynitrides obtained at different temperatures, is constructed. The boundaries between these regions corresponding to coexistence of different phase compositions are identified. The peculiarities of the transitions of the phase compositions of $SiO_xN_y$ films at changing the film stoichiometry and fabrication temperature, together with underlying mechanisms, are discussed with reference to this phase diagram.

A generally good agreement between the theoretical predictions of the $SiO_xN_y$ phase compositions and experimentally determined ones is demonstrated. A poorer agreement for $SiO_xN_y$ films with large Si extent may result from partial excess Si precipitation, not covered by the theoretical model, as well as presence of boundaries between different phase inclusions. Furthermore, reliability of the model predictions decreases at both $x \to 0$ and $y \to 0$, when the film should be regarded rather as Si with small amounts of O and N impurities and not as Si oxynitride composition.

The obtained results extend the fundamental knowledge of the structural properties of Si oxynitride films and mechanisms of their formation as well as may be useful for engineering $SiO_xN_y$ phase composition and, hence, other characteristics for practical applications.

**Acknowledgement**

Authors acknowledge support of their work by the project No. 2025.06/0077 “Technologies for forming passivating coatings for IR radiation photodetectors to increase their detection capability and reliability” of the National Research Foundation of Ukraine.

**Authors’ contributions**

**A.S.:** conceptualization, methodology, validation, investigation, writing – original draft, supervision; **M.B:** methodology, formal analysis, investigation, data curation, writing – review & editing; **M.V.:** methodology, validation, formal analysis, resources, writing – review & editing, **S.V.M.:** writing – review & editing, funding acquisition.

**References**

1. Y. Shi, L. He, F. Guang, L. Li, Z. Xin, R. Liu, A review: Preparation, performance, and

applications of silicon oxynitride film, Micromachines 10 (2019) 552. https://doi.org/10.3390/mi10080552.

2. S. Alexandrova, A. Szekeres, E. Valcheva, M. Anastasescu, H. Stroescu, M. Nicolescu, M. Gartner, Formation of nano-sized silicon oxynitride layers on monocrystalline silicon by nitrogen implantation, Micro 6 (2026) 24. https://doi.org/10.3390/micro6020024.
3. S. Li, M. Li, L. Lan, D. Fu, X. Sun, Z. Gao, Comprehensive investigation on the stability of silicon nitride/oxynitride as thin-film encapsulation layers prepared by plasma-enhanced chemical vapor deposition, ACS Appl. Mater. Interfaces 17 (2025) 10832-10844. https://doi.org/10.1021/acsami.4c20848.
4. N. Hegedüs, C. Balázsi, T. Kolonits, D. Olasz, G. Sáfrán, M. Serényi, K. Balázsi, Investigation of the RF sputtering process and the properties of deposited silicon oxynitride layers under varying reactive gas conditions, Materials 15 (2022) 6313. https://doi.org/10.3390/ma15186313.
5. L. Xu, H. Piao, Z. Liu, C. Cui, D. Yang, Sensitized electroluminescence from erbium doped silicon rich oxynitride light emitting devices, J. Lumin. 235 (2021) 118009. https://doi.org/10.1016/j.jlumin.2021.118009.
6. S. Miyazaki, H. Sakakima, K. Ogawa, S. Izumi, Molecular dynamics study of the effect of composition on elastic properties of silicon oxynitride films, Jap. J. Appl. Phys. 63 (2024) 115502. https://doi.org/10.35848/1347-4065/ad8996.
7. J. Lukeš, V. Kanclíř, J. Václavík, R. Melich, U. Fuchs, K. Žídek, Optically modified second harmonic generation in silicon oxynitride thin films via local layer heating, Sci. Rep. 13 (2023) 8658. https://doi.org/10.1038/s41598-023-35593-8.
8. W.-J. Chen, Y.-C. Liu, Z.-Y. Wang, L. Gu, Y. Shen, H.-P. Ma, Physical and electrical properties of silicon nitride thin films with different nitrogen–oxygen ratios, Nanomaterials 15 (2025) 958. https://doi.org/10.3390/nano15130958.
9. P. Zhang, L. Zhang, F. Lyu, D. Wang, L. Zhang, K. Wu, S. Wang, C. Tang, Luminescent amorphous silicon oxynitride systems: High quantum efficiencies in the visible range, Nanomaterials 13 (2023) 1269. https://doi.org/10.3390/nano13071269.
10. H. Segawa, Y. Osawa, S. Watanabe, S. Machida, K. Katsumata, A. Yasumori, S. Samitsu, K. Deguchi, S. Ohki, N. Ohashi, Investigation of local structures of silicon oxynitride glasses prepared from aerogels, J. Sol-Gel Sci. Technol. 104 (2022) 503-511, https://doi.org/10.1007/s10971-022-05903-z.
11. P. Mota-Santiago, A. Nadzri, F. Kremer, T. Bierschenk, C. E. Canto, M. D. Rodriguez, C. Notthoff, S. Mudie, P. Kluth, Characterisation of silicon oxynitride thin films and their response to swift heavy-ion irradiation, J. Phys. D: Appl. Phys. 55 (2022) 145301. https://doi.org/10.1088/1361-6463/ac45b1.

12. R.-J. Xie, N. Hirosaki, Silicon-based oxynitride and nitride phosphors for white LEDs – A review, Sci. Technol. Adv. Mater. 8 (2007) 588–600. https://doi.org/10.1016/j.stam.2007.08.005.
13. B Li, J. Qin, J. Hong, Y. Gong, L. Sang, K. Shui, T. Xu, X. Chen, J. Zhu, C. Ding, S. Yang, Z. Zhang, Q. Luo, C.-Q. Ma, Inkjet-printed $SiO_xN_y$ barrier film for encapsulation of perovskite solar cells, Small 0 (2026) e74816. https://doi.org/10.1002/smll.74816.
14. R. Aschwanden, R. Köthemann, M. Albert, C. Golla, C. Meier, Optical properties of silicon oxynitride films grown by plasma-enhanced chemical vapor deposition, Thin Solid Films 736 (2021) 138887. https://doi.org/10.1016/j.tsf.2021.138887.
15. T. S. Stokkan, H. Haug, C. K. Tang, E. S. Marstein, J. Gran, Enhanced surface passivation of predictable quantum efficient detectors by silicon nitride and silicon oxynitride/silicon nitride stack, J. Appl. Phys. 124 (2018) 214502. https://doi.org/10.1063/1.5054696.
16. A. Trenti, M. Borghi, S. Biasi, M. Ghulinyan, F. Ramiro-Manzano, G. Pucker, L. Pavesi, Thermo-optic coefficient and nonlinear refractive index of silicon oxynitride waveguides, AIP Adv. 8 (2018) 025311. https://doi.org/10.1063/1.5018016.
17. D. Chen, S. Huang, L. He, Effect of oxygen concentration on resistive switching behavior in silicon oxynitride film, J. Semicond. 38 (2017) 043002. https://doi.org/10.1088/1674-4926/38/4/043002.
18. S. Heo, J. Iee, S. H. Kim, D.-J. Yun, J.-B. Park, K. Kim, N. Kim, Y. Kim, D. Lee, K.-S. Kim, H. J. Kang, Device performance enhancement via a Si-rich silicon oxynitride buffer layer for the organic photodetecting device, Sci. Rep. 7 (2017) 1516. https://doi.org/10.1038/s41598-017-01653-z.
19. M. Perani, N. Brinkmann, A. Hammud, D. Cavalcoli, B. Terheiden, Nanocrystal formation in silicon oxy-nitride films for photovoltaic applications: Optical and electrical properties, The J. Phys. Chem. C 119 (2015) 13907-13914. https://doi.org/10.1021/acs.jpcc.5b02286.
20. C. C. Liu, L. S. Chang, Gas permeation properties of silicon oxynitride thin films deposited on polyether sulfone by radio frequency magnetron reactive sputtering in various $N_2$ contents in atmosphere, Thin Solid Films 594 (2015) 35-39. https://doi.org/10.1016/j.tsf.2015.10.004.
21. Z. B. Zhang, Z. H. Shao, Y. M. Luo, P. Y. An, M. Y. Zhang, C. H. Xu, Hydrophobic, transparent and hard silicon oxynitride coating from perhydropolysilazane, Polymer Int. 64 (2015) 971-978. https://doi.org/10.1002/pi.4871.
22. P. Zhang, K. Chen, Z. Lin, H. Dong, W. Li, J. Xu, X. Huang, The role of N-Si-O bonding configurations in tunable photoluminescence of oxygenated amorphous silicon nitride films, Appl. Phys. Lett. 106 (2015) 231103. https://doi.org/10.1063/1.4922465.
23. A. Zelenina, A. Sarikov, S. Gutsch, N. Zakharov, P. Werner, A. Reichert, C. Weiss, M. Zacharias, Formation of size-controlled and luminescent Si nanocrystals from $SiO_xN_y/Si_3N_4$ hetero-

superlattices, J. Appl. Phys. 117 (2015) 175303. https://doi.org/10.1063/1.4919603.

24. A. Ilyas, N. V. Lavrik, H. K. Kim, P. Aswath, V. Varanasi, Silicon oxynitride overlays on bone-implant systems enhance osteogenesis and biomineralization via surface nitrogen incorporation, ACS Appl. Mater. Interfaces, 7 (2015) 18135-18145, https://doi.org/10.1021/acsami.5b03319.
25. V. G. Varanasi, A. Ilyas, M. F. Velten, A. Shah, W. A. Lanford, P. B. Aswath, Role of hydrogen and nitrogen on the surface chemical structure of bioactive amorphous silicon oxynitride films, J. Phys. Chem. B 121 (2017) 8991-9005. https://doi.org/10.1021/acs.jpcb.7b05885.
26. N. Ahuja, K. Awad, S. Yang, H. Dong, A. Mikos, P. Aswath, S. Young, M. Brotto, V. Varanasi, $SiON_x$ coatings regulate MSC response and reduce bacterial growth, Antioxidants 13 (2024) 189. https://doi.org/10.3390/antiox13020189.
27. S. J. Choo, B. C. Lee, S. M. Lee, J. H. Park, H. J. Shin, Optimization of silicon oxynitrides by plasma-enhanced chemical vapor deposition for an interferometric biosensor, J. Micromech. Microeng. 19 (2009) 095007, https://doi.org/10.1088/0960-1317/19/9/095007.
28. J. P. Lafleur, A. Jönsson, S. Senkbeil, J. P. Kutter, Recent advances in lab-on-a-chip for biosensing applications, Biosens. Bioelectron. 76 (2016) 213–233, https://doi.org/10.1016/j.bios.2015.08.003.
29. K. Saunders, T. Herffurth, S. Bublitz, C. Mühlig, A.-S. Munser, E. Herguedas, A. Redondo-Cubero, D. Gibson, C. Clark, C. García Nuñez, S. Schröder, Optical losses in $SiN_x$ and $SiO_xN_y$ coatings deposited by plasma-enhanced chemical vapor deposition for gravitational wave detectors, Appl. Optics. 65 (2026) A187-A200. https://doi.org/10.1364/ao.578011.
30. D. Criado, A. Zúñiga, I. Pereyra, Structural and morphological studies on $SiO_xN_y$ thin films, J. Non Cryst. Solids 354 (2008) 2809-2815. https://doi.org/10.1016/j.jnoncrysol.2007.09.063.
31. S. Kohli, J. A. Theil, P. C. Dippo, R. K. Ahrenkiel, C. D. Rithner, P. K. Dorhout, Chemical, optical, vibrational and luminescent properties of hydrogenated silicon rich oxynitride films, Thin Solid Films 473 (2005) 89-97. https://doi.org/10.1016/j.tsf.2004.07.054.
32. V. Naseka, I. Nasieka, M. Voitovych, A. Sarikov, I. Lisovskyy, V. Strelchuk, Photoluminescence and Raman scattering behavior of Si rich silicon oxynitride films annealed at different temperatures, Solid State Phenom. 205-206 (2014) 492–496. https://doi.org/10.4028/www.scientific.net/SSP.205-206.492.
33. M. V. Voitovych, A. V. Sarikov, V. O. Yukhymchuk, V. V. Voitovych, M. O. Semenenko, Identification of formation of amorphous Si phase in $SiO_xN_y$ films produced by plasma enhanced chemical vapor deposition, Optical Mater. 174 (2026) 117895. https://doi.org/10.1016/j.optmat.2026.117895.
34. A. Sarikov, M. Voitovych, I. Lisovskyy, V. Naseka, A. Hartel, D. Hiller, S. Gutsch, M. Zacharias, Characteristics of hydrogen effusion from the Si–H bonds in Si rich silicon oxynitride films for nanocrystalline silicon based photovoltaic applications, Adv. Mater. Res. 854 (2014) 69-74. https://doi.org/10.4028/www.scientific.net/AMR.854.69.

35. B. Fischer, A. Lambertz, M. Nuys, W. Beyer, W. Duan, K. Bittkau, K. Ding, U. Rau, Insights into the Si-H bonding configuration at the amorphous/crystalline silicon interface of silicon heterojunction solar cells by Raman and FTIR spectroscopy, Adv. Mater. 35 (2023) 2306351. https://doi.org/10.1002/adma.202306351.
36. V. A. Gritsenko, J. B. Xu, R. W. M. Kwok, Y. H. Ng, I. H. Wilson, Short range order and the nature of defects and traps in amorphous silicon oxynitride governed by the Mott rule, Phys. Rev. Lett. 81 (1998) 1054-1057. https://doi.org/10.1103/PhysRevLett.81.1054.
37. V. A. Gritsenko, R. W. M. Kwok, H. Wong, J. B. Xu, Short-range order in nonstoichiometric amorphous silicon oxynitride and silicon-rich nitride, J. Non Cryst. Solids 297 (2002) 96–101, https://doi.org/10.1016/S0022-3093(01)00910-3.
38. P. Cova, S. Poulin, O. Grenier, R. A. Masut, A method for the analysis of multiphase bonding structures in amorphous $SiO_xN_y$ films, J. Appl. Phys. 97 (2005) 073518. https://doi.org/10.1063/1.1881774.
39. F. Rebib, E. Tomasella, E. Bêche, J. Cellier, M. Jacquet, FTIR and XPS investigations of a-$SiO_xN_y$ thin films structure, J. Phys.: Conf. Series 100 (2008) 082034. https://doi.org/10.1088/1742-6596/100/8/082034.
40. I. P. Lisovskyy, M. V. Voitovych, A. V. Sarikov, S. O. Zlobin, A. N. Lukianov, O. S. Oberemok, O. V. Dubikovsky, Infrared study of the structure of silicon oxynitride films produced by plasma enhanced chemical vapor deposition, J. Non-Cryst. Solids 617 (2023) 122502. https://doi.org/10.1016/j.jnoncrysol.2023.122502.
41. L. Scopel, M. C. A. Fantini, M. I. Alayo, I. Pereyra, Structural investigation of Si-rich amorphous silicon oxynitride films, Thin Solid Films 425 (2003) 275-281. https://doi.org/10.1016/S0040-6090(02)01053-2.
42. M. I. Alayo, I. Pereyra, W. L. Scopel, M. C. A. Fantini, On the nitrogen and oxygen incorporation in plasma-enhanced chemical vapor deposition (PECVD) $SiO_xN_y$ films, Thin Solid Films 402 (2002) 154-161. https://doi.org/10.1016/S0040-6090(01)01685-6.
43. T. Hänninen, S. Schmidt, J. Jensen, L. Hultman, H. Högberg, Silicon oxynitride films deposited by reactive high power impulse magnetron sputtering using nitrous oxide as a single-source precursor, J. Vac. Sci. Technol. A 33, (2015) 05E121. https://doi.org/10.1116/1.4927493.
44. A. Sarikov, Thermodynamic theory of phase separation in nonstoichiometric Si oxide films induced by high-temperature anneals, Nanomanufacturing 3 (2023) 293-314. https://doi.org/10.3390/nanomanufacturing3030019.
45. W. Huang, J. Wang, Q. Xu, M. Yang, K. Liu, J. Peng, C. Wang, R. Tu, S. Zhang, Computational

thermodynamic study on CVD of silicon oxynitride films from Si–O–N–H and Si–O–N–H–Cl systems, Ceram. Int. 50 (2024) 13439-13446. https://doi.org/10.1016/j.ceramint.2024.01.256.

46. J. Wang, B. Xu, K. Lee, W. Huang, H. Wang, J. Peng, M. Xu, Machine learning assisted CALPHAD framework for thermodynamic analysis of CVD $SiO_xN_y$ thin films, Calphad 88 (2025) 102806. https://doi.org/10.1016/j.calphad.2025.102806.
47. M. Babiichuk, A. Sarikov, M. Voitovych, V. Voitovych, Phase composition versus stoichiometry and fabrication temperature of Si oxynitride films by thermodynamic modeling, 2025 IEEE 15th Int. Conf. "Nanomaterials: Applications & Properties" (IEEE NAP-2025), Bratislava, Slovakia, Sep. 7-12 (2025) MTFC08. https://doi.org/10.1109/NAP68437.2025.11216220.
48. D. R. Hamann, Energetics of silicon suboxides, Phys. Rev. B 61 (2000) 9899-9901. https://doi.org/10.1103/PhysRevB.61.9899.
49. A. Bongiorno, A. Pasquarello, Validity of the bond-energy picture for the energetics at Si-$SiO_2$ interfaces, Phys. Rev. B 62 (2000) R16326-R16329. https://doi.org/10.1103/PhysRevB.62.R16326.
50. A. Zelenina, A. Sarikov, D. M. Zhigunov, C. Weiss, N. Zakharov, P. Werner, L. López-Conesa, S. Estradé, F. Peiró, S. A. Dyakov, M. Zacharias, Silicon nanocrystals in $SiN_x/SiO_2$ hetero-superlattices: the loss of size control after thermal annealing, J. Appl. Phys. 115 (2014), 244304. https://doi.org/10.1063/1.4884839.
51. A. Sarikov, M. Zacharias, Gibbs free energy and equilibrium states in the Si/Si oxide systems, J. Phys. Condens. Matter 24 (2012), 385403. https://doi.org/10.1088/0953-8984/24/38/385403.
52. A. M. Hartel, D. Hiller, S. Gutsch, P. Löper, S. Estradé, F. Peiró, B. Garrido, M. Zacharias, Formation of size-controlled silicon nanocrystals in PECVD-grown $SiO_xN_y/SiO_2$ superlattices, Thin Solid Films 520 (2011) 121-125. https://doi.org/10.1016/j.tsf.2011.06.084.
53. M. Ribeiro, I. Pereyra, M. I. Alayo, Silicon rich silicon oxynitride films for photoluminescence applications, Thin Solid Films 426 (2003) 200-204. https://doi.org/10.1016/S0040-6090(03)00008-7.
54. M. V. Fock. Resolution of complex spectra into separate bands using the Alentsev method, Trudy Fiz. Inst. Akad. Nauk SSSR 59 (1972) 3-24 (in Russian).
55. I. P. Lisovskii, V. G. Litovchenko, V. B. Lozinskii, H. Flietner, W. Füssel, E. G. Schmidt, IR study of short-range and local order in $SiO_2$ and $SiO_x$ films, J. Non-Cryst. Solids 87 (1995) 91-95. https://doi.org/10.1016/0022-3093(95)00118-2.
56. I. P. Lisovskyy, M. V. Voitovich, A. V. Sarikov, V. G. Litovchenko, A. B. Romanyuk, V. P. Melnyk, I. M. Khatsevich, P. E. Shepeliavyi, Transformation of the structure of silicon oxide during the formation of Si nanoinclusions under thermal annealings, Ukr. J. Phys. 54 (2009) 383-390.
57. I. P. Lisovskii, V. G. Litovchenko, V. B. Lozinskii, G. I. Steblovskii, IR spectroscopic investigation

of $SiO_2$ film structure, Thin Solid Films 213 (1992) 164–169. https://doi.org/10.1016/0040-6090(92)90278-J.

58. A. Sarikov, M. Semenenko, S. Shahan, Percolation threshold in annealed ultrathin $SiO_x$ films by 2D Monte Carlo simulations, CrystEngComm 26 (2024) 2836-2842. https://doi.org/10.1039/d4ce00212a.
59. Ye. Karpov, A. Sarikov, O. Ryzhko, Monte Carlo modeling of the kinetics of separation and morphology of Si phase during thermally stimulated decomposition of nonstoichiometric Si oxide, Func. Mater. 33 (2026) 343-352. http://dx.doi.org/10.15407/fm33.02.343.
60. B. Kaghouche, F. Mansour, C. Molliet, B. Rousset, P. Temple-Boyer, Investigation on optical and physico-chemical properties of LPCVD $SiO_xN_y$ thin films, The European Phys. J. – Appl. Phys. 66 (2014) 20301. https://doi.org/10.1051/epjap.2014130550.
61. G. Beshkov, S. Lei, V. Lazarova, N. Nedev, S. S. Georgiev, IR and Raman absorption spectroscopic studies of APCVD, LPCVD and PECVD thin SiN films, Vacuum 69 (2002) 301-305. https://doi.org/10.1016/S0042-207X(02)00349-4W.